\documentclass[journal]{IEEEtran}

\usepackage[T1]{fontenc}
\usepackage[utf8]{inputenc}
\usepackage{cite}
\usepackage{graphicx}
\usepackage{amsmath,amssymb,bm}
\usepackage{booktabs}
\usepackage{multirow}
\usepackage{threeparttable}
\usepackage{array}
\usepackage[caption=false,font=footnotesize]{subfig}
\usepackage{xcolor}
\usepackage{url}
\usepackage[hidelinks]{hyperref}

\graphicspath{{figures/}}

\newcommand{\secref}[1]{Section~\ref{#1}}
\newcommand{\figref}[1]{Fig.~\ref{#1}}
\newcommand{\tabref}[1]{Table~\ref{#1}}
\newcommand{\eqnref}[1]{(\ref{#1})}
\newcommand{\lead}[1]{\smallskip\noindent\textbf{#1}\;}
\newcommand{\sysname}{MA-HGAT}

\newcommand{\up}{$\uparrow$}

\newcommand{\best}[1]{\textbf{#1}}
\newcommand{\second}[1]{\underline{#1}}

\begin{document}

\title{Detecting Logic Vulnerabilities Across the Contract and Device Layers of Blockchain-Enabled IoT With Multi-Agent Heterogeneous Graph Attention}

\author{
        Minfeng~Qi,
        Jialin~Li,
        Tianqing~Zhu,
        Lefeng~Zhang,
        and~Zhe~Sun%
\thanks{M.~Qi, J.~Li, T.~Zhu, and L.~Zhang are with the Faculty of Data Science, City University of Macau, Macau, China (e-mail: mfqi@cityu.edu.mo; d24091110649@cityu.edu.mo; tqzhu@cityu.edu.mo; lfzhang@cityu.edu.mo).}%
\thanks{Z.~Sun is with Guangzhou University, Guangzhou, China (e-mail: sunzhe@gzhu.edu.cn).}%
\thanks{Corresponding author: Tianqing Zhu.}%
\thanks{A preliminary version of this article, covering the contract layer only, appeared in the International Symposium on Cyberspace Safety and Security (CSS)~\cite{li2026mahgat_css}. This version adds the system-level formulation for blockchain-enabled IoT (\secref{sec:system}), the task heads (\secref{sec:heads}), the gateway--cloud partitioning (\secref{sec:edgecloud}), the per-graph implementation, and all device-layer and deployment experiments (\secref{sec:iot_experiments}).}}


\maketitle

\begin{abstract}
Blockchain-enabled Internet of Things (IoT) systems integrate smart contracts with embedded devices to support decentralized device management and access control.
Their security therefore depends jointly on the logic of on-chain contracts and off-chain device firmware.
Logic flaws in either layer can violate the same system invariants, such as unauthorized access, improper state changes, or unguarded privileged operations.
Existing approaches rely on contract analysis, firmware analysis, and graph-based vulnerability detection.
However, these methods typically focus on a single layer or artifact and often depend on predefined vulnerability patterns, emulation fidelity, or homogeneous representations that obscure security-relevant component roles.
They also lack a unified architecture that supports different security tasks while remaining deployable on resource-constrained gateways.
To address these limitations, we extend MA-HGAT into a cross-layer multi-agent heterogeneous graph attention framework that models contracts, firmware artifacts, device fleets, and transaction streams with a unified four-role, nine-relation schema.
Role-aligned agents exchange heterogeneous evidence through cross-attention, while graph-, link-, and node-level heads support multiple detection tasks and a role-based gateway--cloud partition enables lightweight edge inference.
MA-HGAT thus provides a unified and deployable framework for detecting logic vulnerabilities across the contract and device layers of blockchain-enabled IoT systems.
\end{abstract}


\IEEEpeerreviewmaketitle

\section{Introduction}
\label{sec:introduction}

\IEEEPARstart{B}{lockchain-enabled} Internet of Things systems replace a centralized trust anchor with a ledger on which smart contracts register devices, enforce access policies, and validate device data~\cite{novo2018blockchain,reyna2018blockchain,reis2025edgechainguard}. The devices themselves remain embedded systems running vendor firmware, while gateways relay requests, status information, and transactions between devices and the blockchain. This architecture has been adopted in smart grids, supply chains, and industrial monitoring because it reduces dependence on a single trusted server and provides a shared audit trail. However, system correctness no longer depends on one software boundary: it depends jointly on the contract logic that governs devices and the firmware logic that controls them. A design mistake at either side can therefore violate the same system policy, for example by admitting an unauthorized device or allowing an attacker to reconfigure an authorized one~\cite{hacken2025,panews2025,antonakakis2017mirai,enisa2024threat}.

Existing security techniques address this problem from several directions. Smart contract analyzers use static analysis, symbolic execution, fuzzing, or semantic reasoning to identify vulnerable contract behavior~\cite{feist2019slither,mythril_tool,luu2016oyente,jiang2018contractfuzzer,kframework2020,sun2024gptscan,ref:smartllamadpo2025}. Firmware analyzers recover authentication and input-validation weaknesses from binaries, emulate devices for testing, or search related firmware images for previously disclosed vulnerabilities~\cite{shoshitaishvili2015firmalice,redini2020karonte,chen2021satc,zheng2019firmafl,feng2020p2im,kim2020firmae,feng2016genius,xu2017gemini,xiao2024firmrec,cheng2025firmvullinker}. Learning-based approaches further model program structure from labeled examples~\cite{sun2025bhgnn,liu2025compsac,duan2019vulsniper}. These approaches are effective within their intended settings, but they remain fragmented: rule-driven methods depend on predefined vulnerability patterns, dynamic methods depend on faithful execution environments, and learning-based methods often simplify security-relevant components into a uniform representation. More importantly, these methods usually analyze either contracts or firmware in isolation and are not designed around the gateway where the two sides of the system meet.

\begin{figure}[t]
    \centering
    \includegraphics[width=\columnwidth]{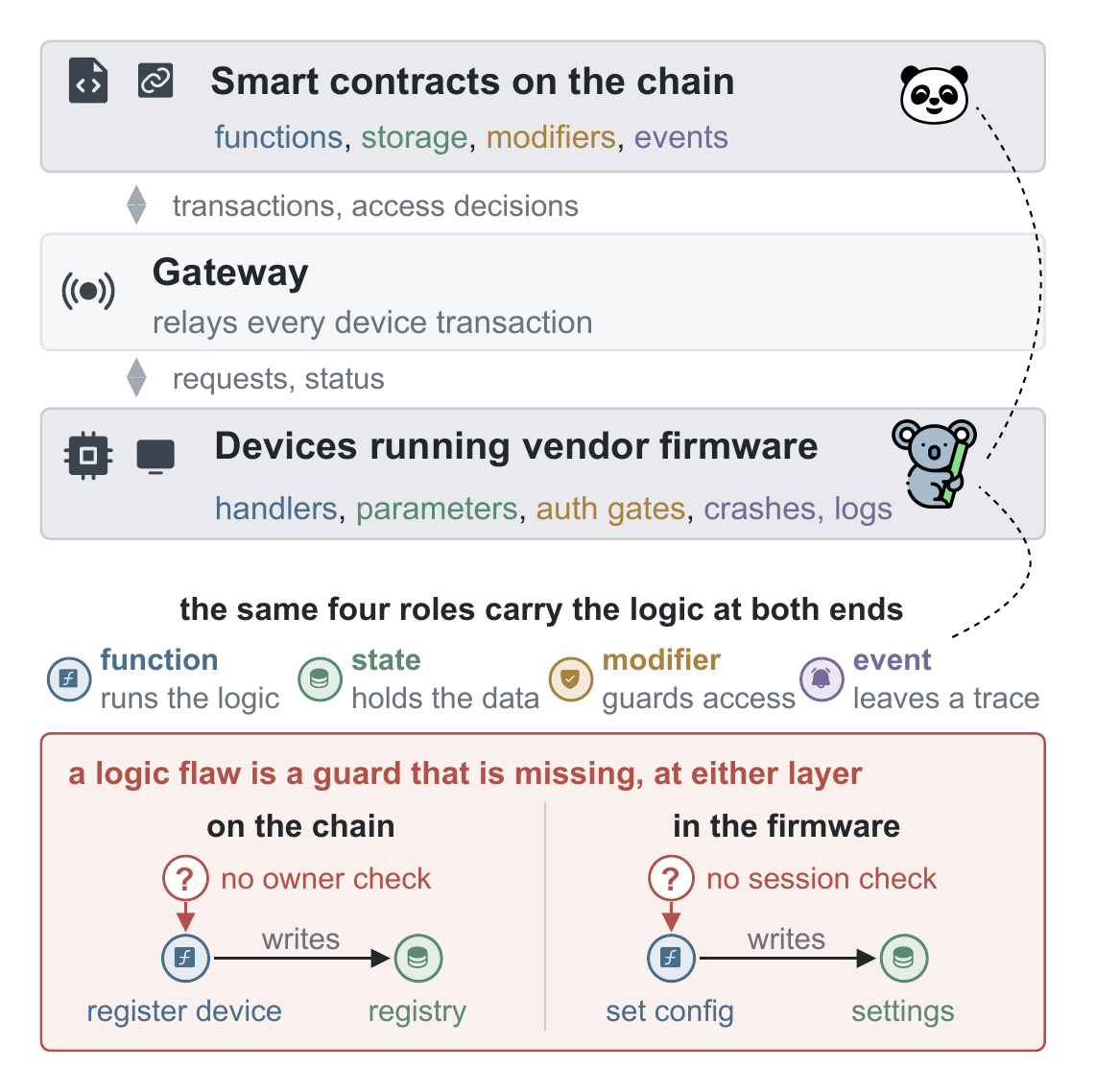}
    \caption{A blockchain-enabled Internet of Things system in which smart contracts and device firmware jointly enforce system logic. The same four security roles appear at both sides: components perform actions, hold state, guard access, and leave observable traces.}
    \label{fig:system}
\end{figure}

The resulting gap is not simply the absence of another vulnerability detector, but the absence of a common basis for reasoning about security across the whole system. In practice, an operator must answer four related questions: whether a contract contains unsafe logic, what type of weakness is described by a device vulnerability report, which deployed devices are affected by a newly disclosed weakness, and whether the behavior of a device appears malicious. These questions are currently handled by separate tools even though they concern the same security policies. \figref{fig:system} illustrates why this separation is problematic. At the contract side, a device-registration operation may update the registry without checking ownership; at the firmware side, a configuration operation may update device settings without checking the current session. Although the concrete software components are different, both failures have the same underlying form: an operation changes protected state without the required guard. The central problem is therefore whether these cross-layer failures can be described and detected through a common security view, while still supporting the different decisions required by operators.

Addressing this problem introduces three main challenges. First, the evidence is structurally diverse: contracts, firmware requests, deployed device images, and transaction streams contain different kinds of components, yet the security meaning often lies in how those components interact rather than in any component alone. Treating all components as equivalent removes distinctions such as whether an operation is guarded, while analyzing each component type independently loses the relationships that expose the flaw. Second, the required decisions have different forms. Contract auditing and vulnerability-report analysis require judgments about an entire artifact, fleet analysis requires identifying which devices are related to a disclosed weakness, and online monitoring requires a decision for each observed action. A common approach must therefore preserve the same security reasoning while supporting these different decision forms. Third, deployment introduces a physical constraint: gateways observe device activity first but have limited computation, memory, and connectivity, whereas the broader context needed to interpret that activity is maintained in the cloud.

Our design is motivated by two observations that connect these challenges. First, despite their different implementations, security-relevant components across the contract and device layers repeatedly play the same four roles: some perform actions, some hold state, some guard access, and some record what happened. A logic vulnerability can therefore be viewed as an incorrect or missing relationship among these roles rather than as a layer-specific code pattern. This view explains why the two examples in \figref{fig:system} are security-equivalent even though one occurs in a smart contract and the other in firmware. Second, among these roles, device actions are the information that a gateway observes directly, while broader state, identity, and historical context can remain in the cloud. These observations suggest that cross-layer detection should be organized around shared security roles rather than software layers, and that deployment should be divided according to what each location naturally observes.

Based on this rationale, we build one framework that represents contracts, firmware evidence, device fleets, and device--contract activity using the same role-based security structure. The framework preserves the distinctions among actions, state, guards, and observable traces while allowing evidence from these roles to inform one another, addressing the first challenge. It then uses the resulting shared representation to support whole-artifact assessment, fleet-level association, and event-level monitoring, addressing the second challenge without requiring a separate reasoning pipeline for each task. Finally, only the part that processes locally observed device actions is placed on the gateway, while the broader system context remains in the cloud, addressing the deployment challenge while avoiding the transmission of raw device requests. We realize this design in our multi-agent heterogeneous graph attention framework.

This article extends the conference version of our framework~\cite{li2026mahgat_css}. The main contributions are as follows:
\begin{itemize}
    \item \textbf{Cross-layer problem formulation.} We formulate logic vulnerability detection in blockchain-enabled Internet of Things systems as a unified problem spanning smart contracts, device firmware, deployed device fleets, and device--contract activity. We show that the same four security roles and their relationships provide a common basis for describing logic flaws across these settings (\secref{sec:system}, \tabref{tab:schema_map}).

    \item \textbf{One framework for four security decisions.} We extend the original contract-oriented framework so that one shared representation supports contract auditing, vulnerability-report analysis, fleet-level vulnerability linking, and transaction monitoring rather than requiring a separate model for each task (\secref{sec:architecture}).

    \item \textbf{Gateway--cloud deployment aligned with system roles.} We divide inference according to the information naturally available at the gateway and in the cloud. Only the component that processes executed device actions runs at the gateway, reducing the local model footprint and replacing raw requests with fixed-size representations while preserving the overall decision process (\secref{sec:edgecloud}, \secref{sec:iot_edgecloud}).

    \item \textbf{Evaluation across contract and device layers.} We evaluate the framework on contract vulnerability detection, device vulnerability-report analysis, and firmware vulnerability linking, together with a gateway--cloud deployment study. The evaluation further includes component ablations and an unseen-vendor setting to identify both the source of cross-layer transfer and the generalization boundary of the learned representation (\secref{sec:contract_experiments}, \secref{sec:iot_experiments}).
\end{itemize}

\section{Related Work}
\label{sec:related}

\subsection{Security of Blockchain-Enabled IoT}

Blockchains entered IoT architectures as a decentralized substitute for the trusted server that registers devices, mediates access, and audits data~\cite{christidis2016blockchains,dorri2017blockchain}. Novo~\cite{novo2018blockchain} showed that access management for large device fleets scales when contracts enforce the policies and devices reach the chain through management hubs. Surveys catalog the resulting designs for supply chains, energy, healthcare, and industrial monitoring~\cite{reyna2018blockchain,qi2025sok}. Recent work evaluates such architectures on 5G edge deployments and releases simulated device--contract transaction logs for that purpose~\cite{reis2025edgechainguard}. Security analyses of these systems concentrate on the ledger and the protocol, and they treat contract correctness and device integrity as assumptions. This article treats both as objects of analysis. The contract that enforces a policy and the firmware that a registered device runs are both places where a logic mistake breaks the policy. We detect logic vulnerabilities at both.

\subsection{Smart Contract Vulnerability Detection}

Research on smart contract vulnerability detection has followed three paths: static analysis, dynamic analysis, and formal verification~\cite{deng2023smart_contract,sun2024gptscan}. Static analyzers such as Mythril~\cite{mythril_tool}, Slither~\cite{feist2019slither}, and Oyente~\cite{luu2016oyente} combine symbolic execution, control- and data-flow analysis, and pattern-based vulnerability templates. Securify~\cite{tsankov2018securify}, ZEUS~\cite{kalra2018zeus}, Sereum~\cite{rodler2019sereum}, and SmartShield~\cite{zhang2020smartshield} extend this line with compliance patterns, abstract interpretation, run-time monitoring, and automatic hardening. These tools are effective for well-defined semantic vulnerabilities, but they must predefine what to look for. This limits them on business-logic flaws that arise from implicit invariants or cross-function interactions~\cite{li2024sast}. Dynamic approaches include property-based testing with Echidna~\cite{echidna2020}, hybrid symbolic-concrete exploration with Manticore~\cite{manticore_docs2019}, and fuzzers such as ContractFuzzer~\cite{jiang2018contractfuzzer} and Smartian~\cite{choi2021smartian}. They uncover vulnerabilities that appear only under specific input sequences, but they remain bounded by path explosion and by the quality of the property specifications~\cite{kong2025_profitable_fuzzing}. Formal verification with the K framework~\cite{kframework2020}, F$^\star$~\cite{fstar2021}, or the Solidity SMTChecker~\cite{solidity_smtchecker} offers strong guarantees for high-assurance contracts at the price of substantial specification effort. It also assumes that the specification itself is sound, which is exactly what a logic vulnerability violates. LLM-based auditors such as GPTScan~\cite{sun2024gptscan}, Smart-LLaMA-DPO~\cite{ref:smartllamadpo2025}, and VulnHunt-GPT~\cite{boi2024_vulnhunt_gpt} add semantic reasoning over code and documentation~\cite{li2025blockchain,jiao2024survey}. However, they process contracts as sequential text and remain sensitive to prompt design and to hallucinated findings.

\subsection{IoT Firmware and Device Security Analysis}

Large-scale studies of embedded firmware established early that IoT devices ship with weak authentication, hard-coded credentials, and insecure management interfaces at scale~\cite{costin2014firmware,antonakakis2017mirai}. Surveys organize the resulting detection literature into static, dynamic, and hybrid categories~\cite{qasem2021survey}. Static approaches recover authentication-bypass conditions from binaries (Firmalice~\cite{shoshitaishvili2015firmalice}) or track tainted data across the multiple binaries that implement one service (Karonte~\cite{redini2020karonte}). Others use keywords shared between front-end and back-end code to reduce taint sources (SaTC~\cite{chen2021satc}) or reason about protocol state in bare-metal images (FirmXRay~\cite{wen2020firmxray}). Tools such as BinAbsInspector~\cite{binabsinspector2022} package abstract interpretation for stripped binaries. Dynamic analysis depends on emulation. AVATAR~\cite{zaddach2014avatar} and Avatar$^2$~\cite{muench2018avatar2} coordinate hardware-in-the-loop execution, and P2IM~\cite{feng2020p2im} models peripheral interfaces automatically. FirmAE~\cite{kim2020firmae} scales full-system emulation, and FIRM-AFL~\cite{zheng2019firmafl} makes greybox fuzzing of emulated firmware practical. Fidelity and crash observability, however, remain fundamental obstacles~\cite{muench2018corrupt}. A separate line targets \emph{recurring} vulnerabilities. Vendors reuse code across product families, so a vulnerability found in one image usually affects many others. Genius~\cite{feng2016genius} and Gemini~\cite{xu2017gemini} search for vulnerable functions through graph-based and neural binary similarity, and FirmUp~\cite{david2018firmup} matches procedures across stripped firmware. FirmRec~\cite{xiao2024firmrec} combines vulnerability-specific signatures with symbolic reasoning. FirmVulLinker~\cite{cheng2025firmvullinker} profiles whole images along five dimensions to link homologous vulnerabilities, and it releases the labeled fleet that we use in \secref{sec:iot_experiments}. Most recently, LLM agents assisted by fuzzing (FirmAgent~\cite{ji2026firmagent}) discover vulnerabilities in real firmware. Surveys of agentic AI for IoT cybersecurity~\cite{nageshwaran2026agentic} identify deployment topology (edge, fog, cloud) and latency at the edge as open problems. Our work complements all of these. It does not replace binary analysis or emulation. Instead, it learns over the heterogeneous relations that they expose and applies the same representation to the contracts that govern the devices. It can also be partitioned between gateways and the cloud.

\subsection{Graph Neural Networks for Code and Security Analysis}

GNNs have become a standard tool for program analysis because they model control, data, and structural dependencies directly~\cite{allamanis2018learning,raju2025elegant}. Researchers have applied graph convolutional and attention networks~\cite{hamilton2017graphsage,velickovic2018gat} to abstract syntax trees, control-flow graphs, and data-flow graphs for bug detection, code classification, and vulnerability identification. Examples include attention-based localization of fine-grained vulnerabilities~\cite{duan2019vulsniper} and explainable reentrancy localization in contracts~\cite{liu2025compsac}. In the IoT domain, graph embeddings form the basis of cross-architecture binary similarity for firmware bug search~\cite{feng2016genius,xu2017gemini}. Heterogeneous GNNs such as HAN~\cite{wang2019han} and R-GCN~\cite{schlichtkrull2018rgcn} distinguish node and relation types through hierarchical attention or relation-specific transformations. Recent contract detectors adopt heterogeneous graphs~\cite{sun2025bhgnn}. These models, however, are domain-agnostic. They neither prioritize state-transition semantics nor separate the reasoning of different component types. Their evaluations rarely test whether the typed representation is actually needed. Our schema is security-oriented by construction, and our evaluation includes homogeneous counterparts trained on the same graphs.

\subsection{Attention, Multi-Agent Reasoning, and Edge--Cloud Inference}

Attention mechanisms let a model weight informative context during representation learning~\cite{vaswani2017attention,alammar2018illustrated}. The KVQ formulation separates what is stored (values) from how relevance is computed (queries against keys). In program analysis, this separation helps to reveal the execution dependencies that uniform aggregation averages away. Multi-agent architectures distribute complex reasoning among specialized agents that capture complementary views of the input~\cite{graph_attention_2025}. In graph learning, agent specialization has been shown to increase representational diversity. Separately, edge intelligence~\cite{zhou2019edge} and split computing~\cite{kang2017neurosurgeon,matsubara2022split} study how to partition a neural network between a resource-constrained device and a server so that only intermediate representations cross the network. Existing partitioning work targets layer boundaries of homogeneous networks. \sysname\ instead exposes a \emph{semantic} boundary, the role whose entities a gateway observes, which we use in \secref{sec:edgecloud}.

\begin{table}[t]
\centering
\caption{Roles and their meaning in the contract graph.}
\label{tab:node_types_contract}
\footnotesize
\begin{tabular}{@{}llp{4.6cm}@{}}
\toprule
\textbf{Role} & \textbf{Symbol} & \textbf{Contract-layer meaning} \\
\midrule
Function       & $V_f$ & Encodes executable logic and interaction behavior \\
State variable & $V_s$ & Maintains persistent contract state \\
Modifier       & $V_m$ & Enforces constraints and access-control policies \\
Event          & $V_e$ & Exposes externally observable execution traces \\
\bottomrule
\end{tabular}
\end{table}

\begin{table}[t]
\centering
\caption{Relation types and their meaning in the contract graph.}
\label{tab:edge_types_contract}
\footnotesize
\begin{tabular}{@{}llp{4.2cm}@{}}
\toprule
\textbf{Relation} & \textbf{Symbol} & \textbf{Contract-layer meaning} \\
\midrule
Calls          & $E_{\textsf{calls}}$ & Function $f_1$ invokes function $f_2$ \\
Depends        & $E_{\textsf{depends}}$ & Function $f$ reads state variable $s$ \\
Modifies       & $E_{\textsf{modifies}}$ & Function $f$ writes state variable $s$ \\
Triggers       & $E_{\textsf{triggers}}$ & Function $f$ emits event $e$ \\
Affects        & $E_{\textsf{affects}}$ & Event $e$ is indexed by state variable $s$ \\
Constrained by & $E_{\textsf{constrained\_by}}$ & Function $f$ is guarded by modifier $m$ \\
Returns to     & $E_{\textsf{returns\_to}}$ & Modifier $m$ returns control to function $f$ \\
Invokes        & $E_{\textsf{invokes}}$ & Modifier $m$ calls function $f$ \\
Uses           & $E_{\textsf{uses}}$ & Modifier $m$ accesses state variable $s$ \\
\bottomrule
\end{tabular}
\end{table}

\section{System Model, Threat Model, and Unified Schema}
\label{sec:system}

\subsection{System Model}

\figref{fig:system} shows the blockchain-enabled IoT system that this article targets. It follows the architecture that access-management and data-integrity proposals for IoT have adopted~\cite{christidis2016blockchains,dorri2017blockchain,novo2018blockchain,ali2019applications,reis2025edgechainguard}. Three layers interact.

\lead{Contract layer.} A set of smart contracts on a permissioned or public chain holds the system's shared state and enforces its policies. A device-registry contract records which devices exist and who owns them. An access-control contract decides which principals may read data from, or send commands to, which devices. A data-validation contract checks and time-stamps reported measurements before other parties consume them. Contract functions execute the policies, storage variables hold the registry and the permissions, modifiers guard the functions, and events expose what happened to off-chain observers.

\lead{Gateway layer.} Edge gateways connect devices to the chain. A gateway authenticates the devices behind it, relays their transactions, caches access decisions, and is the first component to observe what a device does. Gateways have limited resources: they are embedded Linux boards or microcontrollers with kilobytes to megabytes of memory. They often reach the cloud through intermittent or metered links.

\lead{Device layer.} The devices run vendor firmware. Beyond sensing and actuation, the firmware exposes management interfaces, typically HTTP, SOAP, or CGI handlers. Through these interfaces, gateways and administrators read status and change configuration. Request handlers execute the device's logic, request parameters and configuration registers hold its state, authentication gates guard the handlers, and crashes, logs, and telemetry expose its behavior. Devices from a few vendors and product families are deployed in large numbers. A weakness in one firmware image therefore usually affects many registered devices.

The layers jointly enforce the invariants that make the system trustworthy. Only registered devices act, only authorized principals change a device's configuration, and every privileged action leaves a trace. Each invariant can be broken from either side. If the registry contract's registration function lacks an ownership check, an attacker registers a rogue device whose data every party then trusts. If a device's firmware exposes a configuration handler that does not validate the session cookie, an attacker reconfigures a legitimately registered device. The device then keeps acting on the chain with valid credentials. The first flaw is in contract logic and the second in firmware logic. Neither is a memory-safety bug: the code does what its author wrote, and the author's design is wrong.

\begin{table*}[t]
\centering
\caption{One schema, four instantiations across the layers of a blockchain-enabled IoT system. Each column lists the entities that populate the four roles and the main relations, the prediction target, and the task head used.}
\label{tab:schema_map}
\scriptsize
\begin{threeparttable}
\begin{tabular}{@{}p{1.7cm}p{2.9cm}p{4.2cm}p{3.5cm}p{4.1cm}@{}}
\toprule
 & \textbf{Contract layer} & \textbf{Device layer: vulnerability reports} & \textbf{Device layer: registered fleet} & \textbf{Gateway layer: transaction stream} \\
 & Task A, DeFiHack / Web3Bugs & Task B, IoTVulBench & Task C, FirmVulLinker & Task D, EdgeChainGuard \\
\midrule
Function $V_f$ & contract functions & request dispatcher, endpoint handler & firmware images & blockchain transactions \\
State $V_s$    & state variables & HTTP query/body parameters & CVE entries & smart contracts (registry, access control, data validation) \\
Modifier $V_m$ & modifiers & authentication gates (cookie, basic auth, SOAPAction, none) & vendors & device identities \\
Event $V_e$    & events & expected service behavior (crash) & device families & time windows \\
\midrule
\textsf{calls}          & $f_1 \rightarrow f_2$ & dispatcher $\rightarrow$ handler & $k$-NN profile similarity between images & consecutive transactions of one device \\
\textsf{depends}/ \textsf{modifies} & reads / writes of $s$ & GET / POST parameter of the handler & image $\leftrightarrow$ CVE (known links) & transaction reads contract; high-gas transaction writes it \\
\textsf{constrained\_by} & $f$ guarded by $m$ & handler guarded by gate & image $\leftrightarrow$ vendor & transaction guarded by device identity \\
\textsf{triggers}/ \textsf{affects} & $f$ emits $e$; $e$ indexed by $s$ & handler $\rightarrow$ crash; crash $\leftrightarrow$ most anomalous parameter & image $\leftrightarrow$ family & transaction $\rightarrow$ window; window $\rightarrow$ active contracts \\
\midrule
Target & contract is vulnerable & vulnerability types (multi-label) & unknown image$\to$CVE links & transaction is malicious \\
Head   & graph classification & graph classification & link prediction & node classification \\
Role here & original results (\secref{sec:contract_experiments}) & accuracy evaluation & accuracy evaluation & deployment workload only\tnote{a} \\
\bottomrule
\end{tabular}
\begin{tablenotes}\scriptsize
\item[a] EdgeChainGuard is a synthetic dataset~\cite{reis2025edgechainguard}: its generator assigns attack subtypes at random, and its binary attack label coincides with transaction failure. We therefore use it only to measure the cost of gateway--cloud inference on a realistic graph shape, not to report detection accuracy.
\end{tablenotes}
\end{threeparttable}
\end{table*}

\subsection{Threat Model}

The adversary can read the chain and the contract code and submit transactions to the contracts. It can send requests to the management interfaces of devices it can reach, and it fully controls any device it has compromised. The adversary cannot break the chain's consensus or cryptography, cannot tamper with gateways, and cannot alter the model or its inputs. The defender is the operator of the system. The operator can audit contract code before deployment (task A) and triage vulnerability reports and proof-of-concept traces that concern the deployed device models (task B). The operator can also determine which registered firmware images share a newly disclosed weakness (task C) and monitor the transaction stream that its gateways relay (task D). Tasks A to C are offline analyses that run in the cloud. Task D is an online analysis whose first stage should run on the gateway.

\subsection{Unified Heterogeneous Schema}
\label{sec:schema}

We represent every artifact that the four tasks examine as a heterogeneous graph
\begin{equation}
G = (V, E, \tau, \rho),
\label{eq:graph}
\end{equation}
where $V$ is the node set, $E$ the edge set, $\tau: V \rightarrow \{f,s,m,e\}$ assigns each node one of four roles, and $\rho: E \rightarrow R$ assigns each edge one of nine relation types. The roles are
\begin{equation}
V = V_f \cup V_s \cup V_m \cup V_e,
\label{eq:nodes}
\end{equation}
with $V_f$ (\emph{function}) the components that execute logic and $V_s$ (\emph{state}) the components that hold or carry state. The set $V_m$ (\emph{modifier}) holds the components that guard execution, and $V_e$ (\emph{event}) holds the components that expose observable traces. The relation types are
\begin{equation}
\begin{aligned}
R = \{&\textsf{calls}, \textsf{depends}, \textsf{modifies}, \textsf{triggers}, \textsf{affects},\\
      &\textsf{constrained\_by}, \textsf{returns\_to}, \textsf{invokes}, \textsf{uses}\},
\end{aligned}
\label{eq:relations}
\end{equation}
which describe execution (\textsf{calls}), data access (\textsf{depends}, \textsf{modifies}, \textsf{uses}), guarding (\textsf{constrained\_by}, \textsf{returns\_to}, \textsf{invokes}), and observability (\textsf{triggers}, \textsf{affects}). Each node $v$ carries a feature vector $\bm{x}_v \in \mathbb{R}^{d_{\tau(v)}}$ whose dimensionality depends on its role. \tabref{tab:node_types_contract} and \tabref{tab:edge_types_contract} give the contract-layer meaning of the roles and relations as introduced in the conference version.

\tabref{tab:schema_map} shows how the same schema is populated at the device and gateway layers. A proof-of-concept request against a device becomes a small graph. In it, the request dispatcher and the endpoint handler execute, the request parameters carry state, the authentication gates guard, and the expected service behavior (a crash) is the trace. A registered fleet becomes one graph. In it, firmware images execute, CVE entries are the state they may carry, and vendors guard, since they determine which code base an image inherits. Product families are the trace through which a weakness becomes visible across devices. A device--contract transaction stream becomes a graph in which transactions execute, the contracts they touch hold state, device identities guard, and time windows are the observable trace. In every case a logic vulnerability corresponds to a wrong or missing relation between roles. One example is a handler that \textsf{modifies} configuration state without being \textsf{constrained\_by} a gate. Another is an image that should \textsf{depend} on a CVE but is not yet known to. A third is a burst of transactions from one identity that \textsf{triggers} a window unlike its history. The graph constructions are given in \secref{sec:iot_graphs}.

\begin{figure*}[!t]
    \centering
    \includegraphics[width=0.98\textwidth]{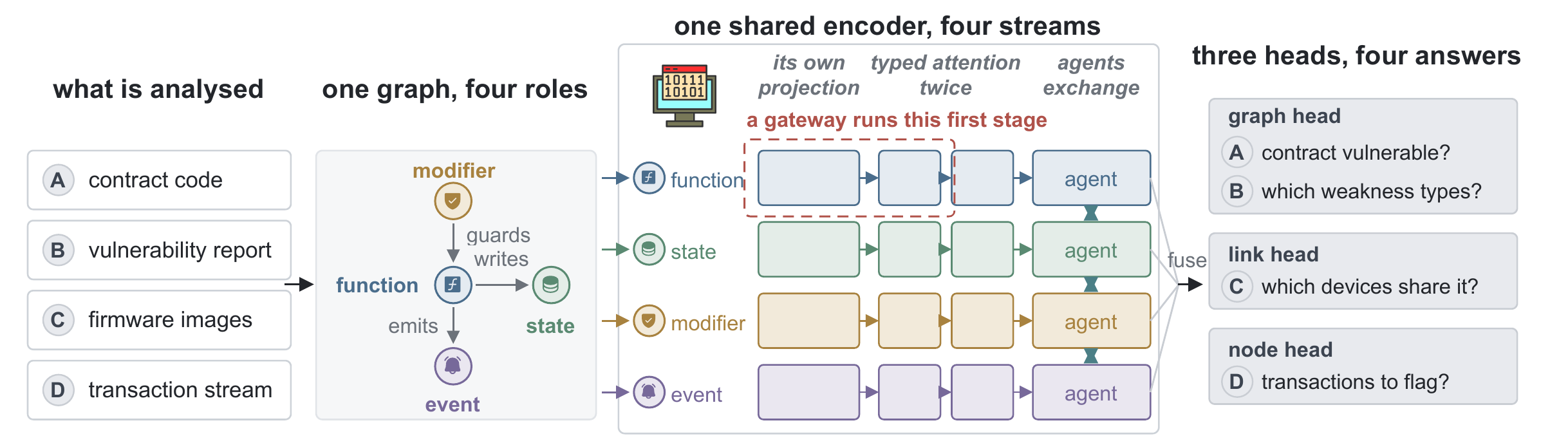}
    \caption{The \sysname\ framework. Each of the four artifacts becomes one graph of the schema of \secref{sec:schema}, in which four roles are joined by named links. The four roles then stay apart as four streams: each has its own projection and its own agent, while the two attention layers are shared. Neighboring agents exchange evidence through cross-attention, the streams are fused, and three heads answer the four questions.}
    \label{fig:framework}
\end{figure*}

\subsection{Learning Objectives}

The encoder produces role-specific node embeddings $\bm{h}_v \in \mathbb{R}^{d}$ for all $v \in V$, and the four tasks attach three kinds of objectives to it.

\lead{Graph classification (tasks A and B).} For a contract or a proof-of-concept trace, the model maps a pooled representation $\bm{h}_G$ to a label vector $\hat{\bm{y}} \in [0,1]^{L}$. Here $L=1$ for binary contract vulnerability. For the multi-label vulnerability types of IoTVulBench (command injection, buffer overflow, denial of service, remote code execution), $L=4$, and a trace may carry several labels.

\lead{Link prediction (task C).} For a fleet graph, the score of a pair $(u, c)$ of an image $u \in V_f$ and a CVE $c \in V_s$ is
\begin{equation}
\mathrm{score}(u,c) = \kappa \cdot \frac{\bm{h}_u^{\top} \bm{h}_c}{\lVert \bm{h}_u \rVert \, \lVert \bm{h}_c \rVert},
\label{eq:link}
\end{equation}
with temperature $\kappa$. For each CVE, the model ranks candidate images by this score to retrieve the registered images that share the weakness.

\lead{Node classification (task D).} For a transaction stream, each function node (transaction) receives a label $\hat{y}_v$ from its embedding.

In all cases the decision depends jointly on the local attributes of individual components and on the typed relations through which state, guards, and observable traces interact. The conference version established this intuition for contracts, and this article carries it to the device and gateway layers.

\section{MA-HGAT Architecture}
\label{sec:architecture}

\figref{fig:framework} gives an overview. The encoder consists of a role-specific feature projection layer and two heterogeneous graph attention layers with relation-specific parameters. These are followed by a state-centered KVQ attention module, a multi-agent decision module with one agent per role, and a gated cross-type fusion layer. Task heads (\secref{sec:heads}) are attached to the encoder output. The description of the encoder follows the conference version. The task heads, the gating in the fusion layer, the link scorer, and the strictly per-graph computation of the attention and fusion stages are new.

\subsection{Feature Projection Layer}

Functions, state entities, modifiers, and events encode fundamentally different roles. A shared linear transformation would therefore hide type-level inductive biases that matter for security reasoning. Each node type $t \in \{f,s,m,e\}$ therefore has its own projection,
\begin{equation}
\bm{h}_t^{(0)} = \mathrm{ReLU}\big(\mathrm{LN}(\bm{W}_t \bm{x}_t + \bm{b}_t)\big),
\label{eq:proj}
\end{equation}
where $\bm{W}_t \in \mathbb{R}^{d \times d_t}$ and $\bm{b}_t \in \mathbb{R}^{d}$ are type-specific. Layer normalization (LN) stabilizes scales across types, and the projection maps all types into a common hidden dimensionality $d$. Dropout follows the activation. At the device and gateway layers, this layer also absorbs the domain differences. A 37-dimensional byte-level firmware profile, a 10-dimensional HTTP-parameter descriptor, and a one-hot vendor indicator all enter the same encoder through their own $\bm{W}_t$.

\subsection{Heterogeneous Graph Attention Layers}

For each relation $r \in R$ we compute relation-aware attention coefficients
\begin{equation}
\begin{aligned}
\alpha_{ij}^{r} &= \frac{\exp(e_{ij}^{r})}{\sum_{k \in \mathcal{N}_i^{r}} \exp(e_{ik}^{r})},\\
e_{ij}^{r} &= \mathrm{LeakyReLU}\big(\bm{a}_r^{\top}[\bm{W}_r \bm{h}_i \oplus \bm{W}_r \bm{h}_j]\big),
\end{aligned}
\label{eq:attn}
\end{equation}
where $\mathcal{N}_i^{r}$ is the set of neighbors of node $i$ under relation $r$, $\bm{W}_r$ is a relation-specific projection, $\bm{a}_r$ the corresponding attention vector, and $\oplus$ concatenation. We aggregate messages across relations,
\begin{equation}
\bm{h}_i^{(l+1)} = \sigma\Big(\sum_{r \in R} \sum_{j \in \mathcal{N}_i^{r}} \alpha_{ij}^{r} \bm{W}_r \bm{h}_j^{(l)}\Big),
\label{eq:agg}
\end{equation}
with multi-head attention inside each relation. We stack two such layers and sum their outputs, so that both one-hop and two-hop relational context are retained,
\begin{equation}
\bm{h}_i = \bm{h}_i^{(1)} + \bm{h}_i^{(2)}.
\label{eq:residual}
\end{equation}
Keeping a separate $\bm{W}_r$ for each of the nine relations lets the model weight edges by their type. For instance, it can weight a \textsf{modifies} edge from an unauthenticated handler to a configuration parameter differently from a \textsf{depends} edge that only reads it.

\subsection{State Attention Module (Key--Value--Query)}

Logic vulnerabilities often arise from abnormal or unintended state transitions~\cite{rodler2019sereum}: inconsistent updates to privileged variables, violations of invariants, or improper sequencing of persistent operations. To emphasize such transitions the encoder applies a KVQ attention module over the state nodes,
\begin{equation}
\mathrm{Attention}(\bm{Q},\bm{K},\bm{V}) = \mathrm{softmax}\Big(\frac{\bm{Q}\bm{K}^{\top}}{\sqrt{d_k}}\Big)\bm{V},
\label{eq:kvq}
\end{equation}
\begin{equation}
\bm{Q} = \bm{W}_Q\tilde{\bm{H}}_s,\; \bm{K} = \bm{W}_K\tilde{\bm{H}}_s,\; \bm{V} = \bm{W}_V\tilde{\bm{H}}_s,\; \tilde{\bm{H}}_s = \bm{H}_s + \bm{P},
\label{eq:kvq2}
\end{equation}
where $\bm{H}_s$ stacks the state-node embeddings and $\bm{P}$ is a learned positional encoding along the ordering of the state nodes in the graph. This ordering is declaration order for contract variables and parameter order within a request. Queries express what the model is searching for, keys encode candidate transition points, and values carry the semantic context of each state entity. The output is added residually to $\bm{H}_s$ with layer normalization. For contracts, this module concentrates on the storage operations that underlie price manipulation, privilege escalation, and atomicity failures. At the device layer, its usefulness depends on whether the state nodes carry rich attributes. We quantify this effect in \secref{sec:iot_ablation}. The attention is computed inside each graph. When several graphs are batched, keys and values of other graphs are masked out, and the positional encoding indexes state nodes within their own graph.

\subsection{Multi-Agent Decision Module}

To separate reasoning perspectives, \sysname\ uses four agents aligned with the node types: the \textbf{Function} agent (F), the \textbf{State} agent (S), the \textbf{Modifier} agent (M), and the \textbf{Event} agent (E). Each agent first refines the embeddings of its own type through a residual feed-forward block,
\begin{equation}
\bm{A}_a = \mathrm{LN}\big(\bm{H}_a + \mathrm{FFN}_a(\bm{H}_a)\big), \quad a \in \{F,S,M,E\}.
\label{eq:agent}
\end{equation}
The agents then exchange information through cross-attention over their per-graph summary tokens $\bar{\bm{a}}_a = \mathrm{mean}(\bm{A}_a)$ (the mean over the nodes of role $a$ in the graph at hand),
\begin{equation}
[\bm{c}_F,\bm{c}_S,\bm{c}_M,\bm{c}_E] = \mathrm{MultiHeadAttention}([\bar{\bm{a}}_F,\bar{\bm{a}}_S,\bar{\bm{a}}_M,\bar{\bm{a}}_E]),
\label{eq:cross}
\end{equation}
and each agent's nodes receive the exchanged context through a learned gate,
\begin{equation}
\bm{g}_a = \sigma\big(\bm{W}_g[\bm{A}_a \oplus \bm{c}_a]\big),\qquad
\bm{A}_a' = \mathrm{LN}\big(\bm{A}_a + \bm{g}_a \odot \bm{c}_a\big).
\label{eq:gate}
\end{equation}
This design encourages complementary reasoning instead of collapsing all evidence into one homogeneous embedding. For example, the Function agent may see an authorization flaw as an anomalous call pattern. The Modifier agent may see it as an unguarded write, and the Event agent as a missing audit trace. The gate lets each agent decide how much of the others' view to absorb.

\subsection{Gated Cross-Type Fusion Layer}

The fusion layer integrates the four agents once more at the graph level. A second multi-head attention block maps the type summaries $\bar{\bm{a}}_a'$ to fused context vectors $\bm{u}_a$. A residual gate then injects these vectors back into every node of the corresponding type,
\begin{equation}
\bm{h}_v^{\mathrm{final}} = \mathrm{LN}\Big(\bm{h}_v + \sigma\big(\bm{W}_u[\bm{h}_v \oplus \bm{u}_{\tau(v)}]\big) \odot \bm{u}_{\tau(v)}\Big).
\label{eq:fusion}
\end{equation}
The conference version used plain concatenation and projection. The gate instead lets global context modulate node-specific embeddings without overwriting them. This property is necessary for the node-level and link-level heads introduced next.

\subsection{Task Heads}
\label{sec:heads}

\lead{Graph classification.} All nodes of all types are pooled with a learned scalar gate $\gamma_v = \sigma(\bm{w}_\gamma^{\top}\bm{h}_v^{\mathrm{final}})$,
\begin{equation}
\bm{h}_G = \frac{\sum_{v \in V} \gamma_v \bm{h}_v^{\mathrm{final}}}{\sum_{v \in V} \gamma_v + \epsilon},\qquad
\hat{\bm{y}} = \bm{W}_2\,\mathrm{ReLU}(\bm{W}_1 \bm{h}_G + \bm{b}_1) + \bm{b}_2,
\label{eq:pool}
\end{equation}
so the evidence for a graph-level decision may come from any node type. The gate is trained end-to-end to select the informative nodes.

\lead{Link prediction.} The scorer of \eqnref{eq:link} is applied to the final embeddings of a function node (firmware image) and a state node (CVE). Training uses binary cross-entropy over known positive links and uniformly sampled negative images (three per positive). Inference ranks all candidate images per CVE.

\lead{Node classification.} A two-layer perceptron maps $\bm{h}_v^{\mathrm{final}}$ of each function node to class logits.

\subsection{Loss Function and Training}

For the binary contract task the conference version combines binary cross-entropy with focal loss~\cite{lin2017focal},
\begin{equation}
\mathcal{L} = \alpha \mathcal{L}_{\mathrm{BCE}} + (1-\alpha)\mathcal{L}_{\mathrm{Focal}},\quad
\mathcal{L}_{\mathrm{Focal}} = -\sum_i (1-p_i)^{\gamma}\log p_i,
\label{eq:loss}
\end{equation}
to emphasize rare but high-impact positives. The multi-label IoT typing task uses per-label binary cross-entropy, the link task uses binary cross-entropy over sampled pairs, and the node task uses class-weighted cross-entropy. We optimize all models with AdamW~\cite{loshchilov2019adamw}, with dropout on the projection, attention, and fusion layers. We use early stopping or a fixed epoch budget as stated per experiment. The implementation is built on PyTorch and the Deep Graph Library~\cite{wang2019dgl}. A single \texttt{MAHGAT} class exposes the three heads through a \texttt{mode} argument and three ablation switches (\texttt{use\_state\_attn}, \texttt{use\_multi\_agent}, \texttt{use\_cross\_fusion}). For efficiency, the graphs of a fold are processed as one batched graph. The state attention, the agent summaries, the cross-attention, and the fusion context are computed with segment masks, so every graph is processed exactly as if it were alone. We verified that the output for a graph is identical up to single-precision rounding (below $2\times10^{-7}$) whether or not it is batched with other graphs.

\subsection{Contract Graph Construction}

For completeness we recall how contract graphs are built in the conference version. Solidity sources are compiled and parsed to obtain abstract syntax trees and control-flow graphs. Functions, state variables, modifiers, and events become nodes. Their attributes include visibility, parameter and return signatures, storage type and initialization, guarded conditions, and emission locations. Call edges are derived from invocation sites, and \textsf{depends} and \textsf{modifies} edges from state reads and writes. The \textsf{constrained\_by}, \textsf{returns\_to}, and \textsf{invokes} edges come from modifier applications and their control transfers. The \textsf{triggers} and \textsf{affects} edges come from event emissions and indexed parameters~\cite{gorski2024_design_pattern,liu2025compsac,choi2021smartian}.

\section{Detection Workflow and Gateway--Cloud Deployment}
\label{sec:workflow}

This section describes the five-tier workflow in which \sysname\ serves the four tasks of \secref{sec:system}. The first four tiers generalize the pipeline of the conference version. The fifth tier is new and covers deployment on gateways, and \figref{fig:framework} marks the split it introduces.

\subsection{Tiers 1--4: Ingestion, Graph Construction, Detection, and Triage}

\lead{Ingestion and normalization.} We normalize raw artifacts into a common record format. These artifacts are Solidity projects from DeFiHack~\cite{defihacklabs} and Web3Bugs~\cite{web3bugs}, PoC traces and advisories from IoTVulBench, firmware images and ground-truth tables from FirmVulLinker, and transaction logs from EdgeChainGuard. For contracts, we resolve compiler versions and dependencies~\cite{crytic_compile} and filter out incomplete sources~\cite{pinheiro2021validation}. For PoC traces, we parse the HTTP request line, headers, query string, and body. For firmware, we profile each image at the byte level. For transaction logs, we sort records by time and accumulate per-device histories causally.

\lead{Heterogeneous graph construction.} We convert each subject into the schema of \secref{sec:schema}. The contract construction is summarized above. The device- and gateway-layer constructions are described in \secref{sec:iot_graphs}. Every construction guarantees that each canonical relation exists, possibly as a placeholder edge, so that graphs of different subjects can be batched.

\lead{Parallel multimodal detection.} Three engines can run in parallel. The first is the \sysname\ detector. The second is a rule-based detector that encodes interpretable structural rules. Examples at the device layer are shell metacharacters in parameters, oversized single-character runs, or unauthenticated access to privileged endpoints. The third is an optional LLM-based semantic analyzer that reasons over documentation and decompiled code~\cite{sun2024gptscan,boi2024_vulnhunt_gpt}. In the conference experiments, the three engines were combined through score normalization, confidence calibration, and cross-engine consistency filtering. In the device-layer experiments of this article, we evaluate \sysname\ on its own and report the rule-based detector as a baseline. This keeps the accuracy attributable to the graph model from being confounded with ensemble effects.

\lead{Dataset-aware triage.} This tier turns the outputs into the artifacts an analyst needs. These are per-type probabilities for PoC traces, a ranked list of candidate homologous images per CVE, and per-transaction alerts with the device and contract they involve. Dataset-specific post-processing is also applied at this tier: proxy and flash-loan patterns for DeFiHack, multi-file dependencies for Web3Bugs, and vendor and family metadata for firmware fleets.

\subsection{Tier 5: Gateway--Cloud Partitioning}
\label{sec:edgecloud}

Stream monitoring (task D) differs from the offline tasks in where the computation may run. Gateways see the transactions and requests of their own devices immediately, but they have little memory and intermittent uplinks. The context that gives a transaction its meaning lives in the cloud: which contracts and devices exist, and what happened in other time windows. The modules of \sysname\ are aligned with roles, and the Function agent is the only module whose inputs a gateway observes. We therefore split the encoder as follows.

\lead{Gateway tier.} Each gateway runs the function-role projection $\bm{W}_f$ and the \textsf{calls}-relation attention of the first layer. That is, it passes messages along the intra-device temporal chain of transactions it observes. Its output is a $d$-dimensional embedding per event, which is uploaded instead of the raw request.

\lead{Cloud tier.} The cloud receives the gateway embeddings as the function-node inputs. It runs the projections of the other three roles and completes the first layer over all relations. It then runs the second layer, the state attention, the multi-agent module, the fusion layer, and the task head. The gateway embedding already summarizes the intra-device temporal chain of each device, so the raw request features never leave the gateway. The model is trained end to end with the split in place, so the additional local pass is accounted for rather than approximated.

This split has three consequences, which we quantify in \secref{sec:iot_edgecloud}. First, the gateway model is small: the function projection plus one relation-specific attention block. Second, the upstream payload is a fixed-size embedding rather than a variable-size request whose content may be sensitive. Third, the cloud latency is independent of the number of gateways, because the gateway work runs in parallel with no coordination. Unlike layer-wise split computing~\cite{kang2017neurosurgeon,matsubara2022split}, the boundary here is semantic. It follows the role that a gateway observes rather than a depth in the network.

\subsection{What the Split Protects and What It Does Not}
\label{sec:split_security}

Moving one stage of the model to the gateway changes what an adversary can reach, so we state what the boundary buys and what it does not.

\lead{Evasion does not become easier.} The gateway computes the function projection and the \textsf{calls} attention over the transactions of one device. An adversary who controls that device and knows the boundary can shape that chain. The chain is one input to the decision among several. The cloud still projects the other three roles, completes the first layer over all nine relations, and runs the second layer, the state attention, the agents, and the head. Those inputs are the contracts, the device identities registered on the chain, and the time windows. A single compromised device does not control them. The model is also trained end to end with the split in place. The gateway stage is part of one function, not a filter in front of it. \secref{sec:iot_edgecloud} confirms that the decisions match cloud-only inference.

\lead{Content minimization is not a privacy guarantee.} The uplink carries a $d$-dimensional embedding instead of the request. That embedding is trained for the detection objective and carries no bound on what it reveals. An adversary who holds the gateway model and observes its outputs may recover properties of the inputs, as the work on collaborative-inference inversion~\cite{he2019inversion_collabinf} and on leakage from embeddings~\cite{song2020embedding_leakage} shows. Our claim is therefore the narrow one: request contents stay on the device side of the uplink, and the cloud never stores them. An operator who needs a stated bound has to add a mechanism such as calibrated noise, and we did not measure what that would cost in accuracy.

\lead{A compromised gateway is not contained.} The threat model assumes that gateways are not tampered with, and that assumption carries weight here. A compromised gateway can upload any embedding for its own devices. Those embeddings enter the shared graph, so the second layer and the fusion carry them to nodes of the other roles and then to other devices. The boundary does not bound this influence. An operator who cannot trust gateways needs per-gateway rate limits or calibration on top of the model.

\lead{Extraction of the gateway model gives one stage, not the decision.} The 20.5~KB that a gateway holds are the function projection and one relation-specific attention block, 2.5\% of the parameters. An adversary who extracts them learns neither the head nor the modules for the other three roles, so the decision function does not follow from the gateway alone. Extraction does give white-box access to the stage that the adversary's own traffic passes through, which is the usual starting point for building evasive inputs.

\subsection{Device- and Gateway-Layer Graph Constructions}
\label{sec:iot_graphs}

\lead{Vulnerability reports (IoTVulBench, task B).} Each CVE in IoTVulBench~\cite{iotvulbench2025} ships a curated advisory and a raw HTTP PoC payload. The advisory (\texttt{detail.yml}) gives the name, description, CVSS score, severity, and tags. The payload reproduces the vulnerability against an emulated router. These routers are the kind of device whose management interface a gateway or administrator talks to in the system of \secref{sec:system}. We parse the payload into a graph as follows. There are two function nodes: the request dispatcher, which carries the method, and the endpoint handler. The handler carries path descriptors such as depth, length, and the presence of \texttt{goform}, \texttt{HNAP}, \texttt{userRpm}, or \texttt{cgi}, together with request-level statistics. There is one state node per query or body parameter. Its features are length, Shannon entropy, density of shell metacharacters, longest single-character run, URL-encoding density, path-traversal markers, and credential-like names. There is one modifier node per authentication gate found in the headers: cookie, basic authorization, SOAPAction, or an explicit \emph{no authentication} node. There is one event node for the expected service behavior (a crash, attributed with the request length). GET parameters attach to the handler through \textsf{depends}, and POST parameters attach through \textsf{modifies}. Gates attach through \textsf{constrained\_by}, \textsf{invokes}, and \textsf{returns\_to}. A cookie gate attaches through \textsf{uses} to the parameter whose name it references, or to the first parameter when none matches. The event node is \textsf{triggered} by the handler and \textsf{affects} the most anomalous parameter. Labels are taken only from the curated advisory text, and features only from the payload. The CVSS score of the advisory is not used as a feature.

\lead{Registered fleet (FirmVulLinker, task C).} FirmVulLinker~\cite{cheng2025firmvullinker} releases 54 known-defective firmware images (TP-Link and D-Link routers, 24 device families) and a ground-truth table of 74 CVEs. Each CVE entry lists a baseline image and the other images it affects (11.3 on average, from 1 to 34). We build one fleet graph. Firmware images are function nodes, each described by a 37-dimensional byte-level profile computed from the raw image bytes. When the image is an archive, we profile its largest member. The profile holds the log size, overall and 4-KB block entropy statistics, a nibble histogram, and printable-string density and length. It also holds the normalized frequency, over the first 8~MB, of 14 tokens such as \texttt{httpd}, \texttt{goform}, \texttt{login}, and \texttt{password}. CVEs are state nodes (disclosure year), vendors are modifier nodes, and device families are event nodes. Images connect to their vendor (\textsf{constrained\_by}), to their family (\textsf{triggers}), and to their three most similar images by cosine similarity of the profiles (\textsf{calls}). For training links only, images also connect to the CVEs known to affect them (\textsf{depends}).

\lead{Device--contract transaction stream (EdgeChainGuard, task D).} EdgeChainGuard~\cite{reis2025edgechainguard} provides 500 blockchain-mediated IoT transactions from 50 devices against three contracts (access control, device registry, and data validation). This is exactly the traffic that the gateway layer of \secref{sec:system} relays. Transactions are function nodes with causal features only. These features are the normalized gas fee, time of day, contract indicator, log inter-arrival time, burst count, and the prior failure rate and history length of the device. Contracts are state nodes, devices are modifier nodes, and twenty time windows are event nodes. Consecutive transactions of one device are chained by \textsf{calls}. A transaction \textsf{depends} on its contract and \textsf{modifies} it when its gas fee is in the top quartile. Devices \textsf{invoke} and \textsf{constrain} their transactions and \textsf{use} the contracts they touch. Windows are \textsf{triggered} by transactions and \textsf{affect} the contracts active in them. As noted in \tabref{tab:schema_map}, this graph is used only as a deployment workload.

\section{Contract-Layer Results}
\label{sec:contract_experiments}

This section summarizes the contract-layer evaluation (task A) of the conference version~\cite{li2026mahgat_css}. All numbers in this section are those reported there. The datasets are DeFi contracts rather than device-registry contracts, because no labeled corpus of IoT-specific contracts exists. The defect classes they cover are access-control violations, state-update ordering, and arithmetic logic. These are the classes that registry, access-control, and validation contracts are exposed to.

\subsection{Setup}

The conference version used two datasets with complementary characteristics. \textbf{DeFiHack}~\cite{defihacklabs} contains 663 contracts with manually verified logic-vulnerability annotations. These cover DeFi attack scenarios such as price manipulation, flash-loan attacks, reentrancy exploits, and state-inconsistency vulnerabilities. \textbf{Web3Bugs}~\cite{web3bugs} is a corpus of exploitable smart contract bugs. We use the 72-project subset released with the \textsf{GPTScan} evaluation~\cite{web3bugs_gptscan_subset}, which spans access-control weaknesses, arithmetic errors, and logic design flaws. The contracts were compiled with \texttt{solc} 0.8.19 and converted into heterogeneous graphs. Each dataset was split 70/15/15 into training, validation, and test sets, with no contract shared across splits. The model used hidden size 256, four attention heads, dropout 0.2, AdamW with learning rate $10^{-3}$, weight decay 0.01, batch size 32, and early stopping with patience 10. The reported metrics were accuracy, precision, recall, and F1-score.

\subsection{Main Results and Baseline Comparison}

\tabref{tab:contract_main} summarizes the main results. \sysname\ combined with the DeepSeek-R1 semantic analyzer achieved 87.04\% accuracy, 86.55\% precision, 83.51\% recall, and 85.35\% F1 on DeFiHack. On Web3Bugs it achieved 87.45\% accuracy, 88.97\% precision, 85.54\% recall, and 88.42\% F1. Compared with symbolic tools (Slither~\cite{feist2019slither}, Mythril~\cite{mythril_tool}, Smartian~\cite{choi2021smartian}) and fuzzing (ContractFuzzer~\cite{jiang2018contractfuzzer}), the framework yielded consistently higher recall and F1. Relative to the two learning-based baselines of \tabref{tab:contract_main}, it improved F1 by 4.08 points over the stronger one.

\begin{table}[t]
\centering
\caption{Main detection results on DeFiHack as reported in the conference version~\cite{li2026mahgat_css} (\%).}
\label{tab:contract_main}
\footnotesize
\begin{tabular}{@{}lcccc@{}}
\toprule
\textbf{Method} & \textbf{Acc.} & \textbf{Prec.} & \textbf{Rec.} & \textbf{F1} \\
\midrule
Slither~\cite{feist2019slither}            & 72.34 & 75.21 & 68.45 & 71.67 \\
Mythril~\cite{mythril_tool}                & 68.92 & 71.34 & 65.78 & 68.45 \\
Smartian~\cite{choi2021smartian}           & 76.89 & 78.23 & 74.56 & 76.35 \\
ContractFuzzer~\cite{jiang2018contractfuzzer} & 74.56 & 76.12 & 72.34 & 74.19 \\
\midrule
GCN+VulDetector                            & 79.23 & 80.67 & 77.45 & 79.03 \\
ReGNN                                      & 81.56 & 82.34 & 80.23 & 81.27 \\
\midrule
\sysname+DeepSeek-R1                       & \best{87.04} & \best{86.55} & \best{83.51} & \best{85.35} \\
\bottomrule
\end{tabular}
\end{table}

\subsection{Comparison with LLM-Based Detectors}

\begin{figure}[t]
    \centering
    \subfloat[DeFiHack ($n=663$)]{\includegraphics[width=0.49\columnwidth]{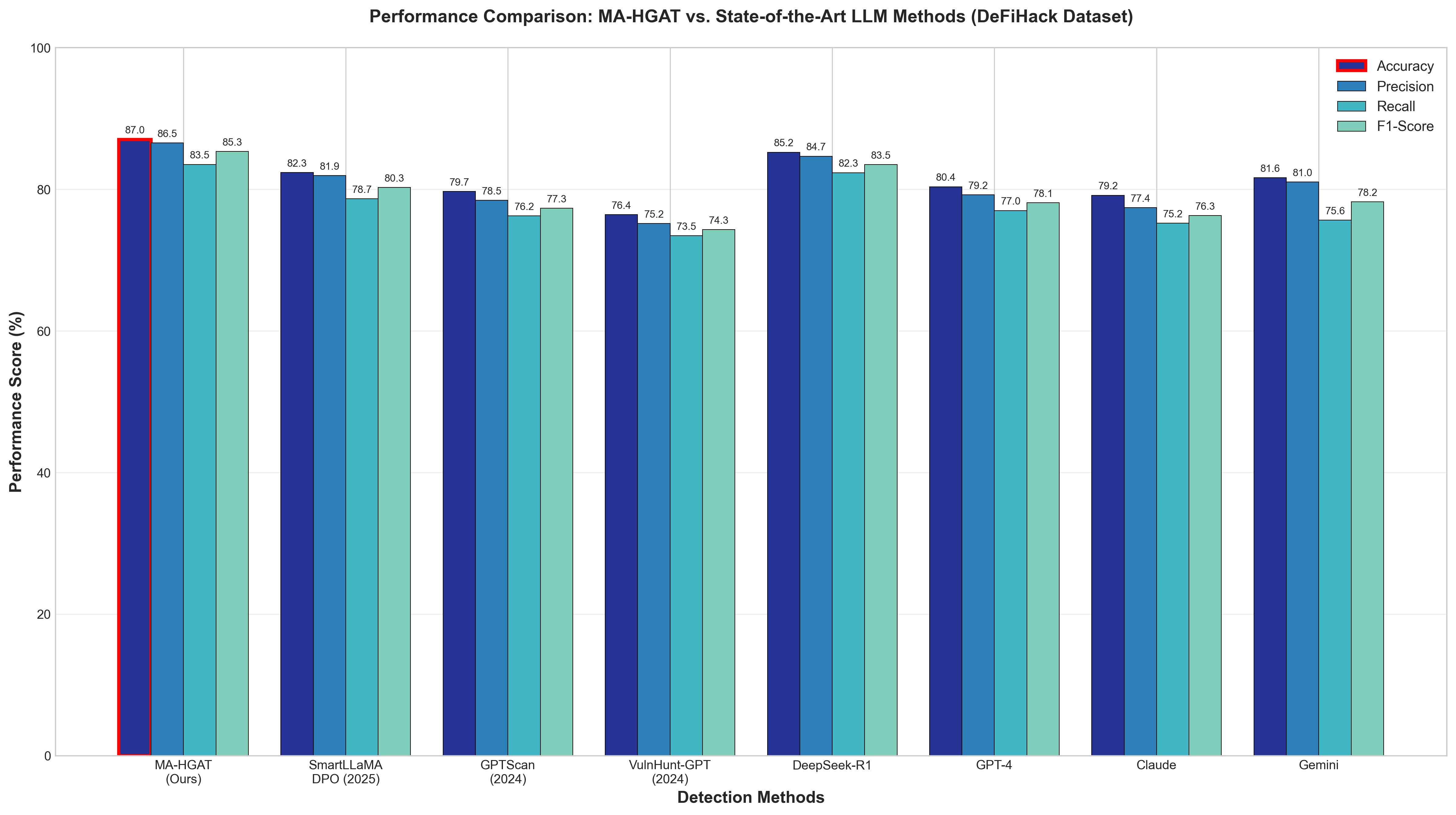}}\hfill
    \subfloat[Web3Bugs ($n=72$)]{\includegraphics[width=0.49\columnwidth]{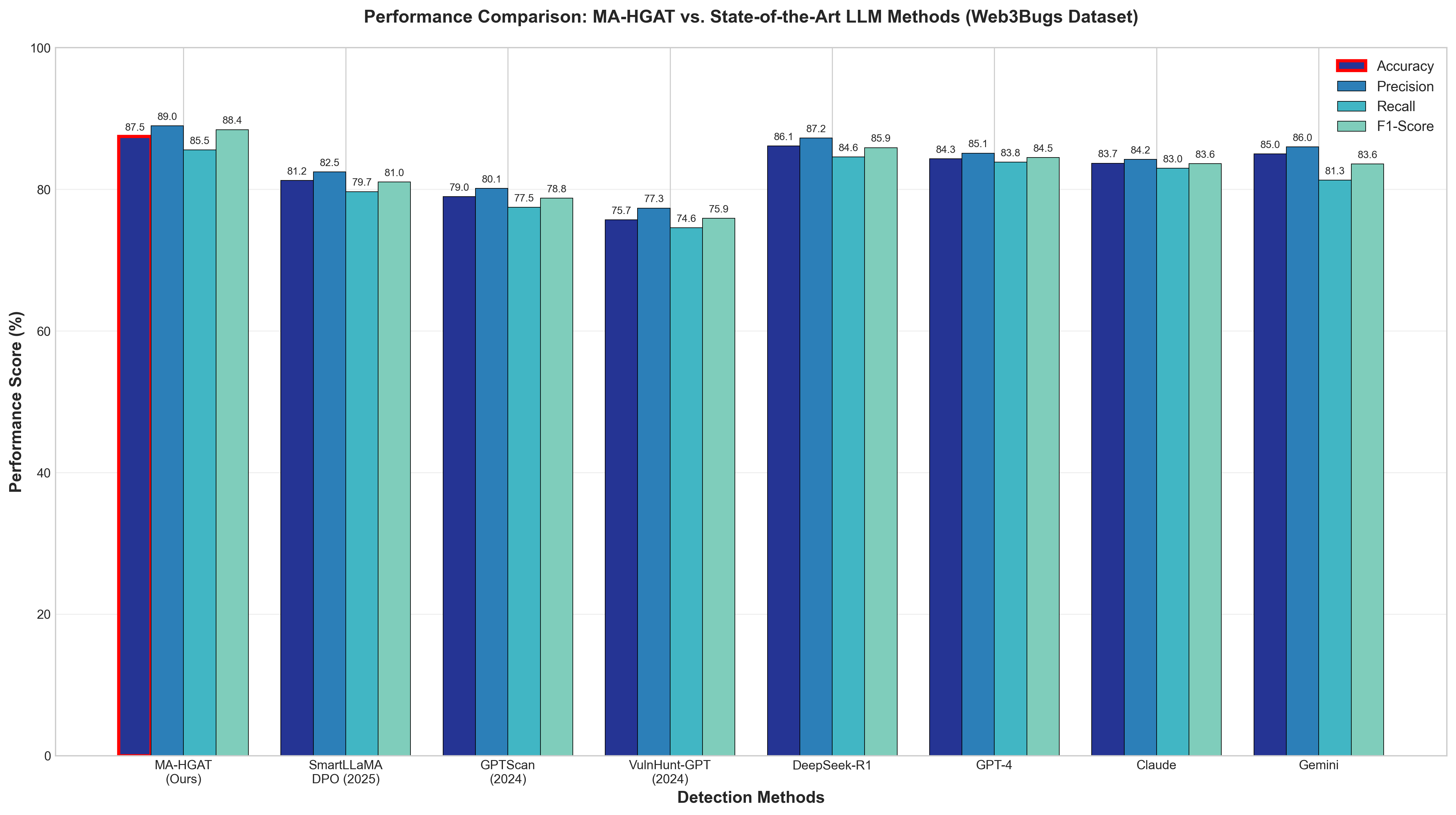}}
    \caption{Comparison with LLM-based detectors on the two contract datasets.}
    \label{fig:contract_llm}
\end{figure}

\figref{fig:contract_llm} compares \sysname\ with LLM-based detectors. On DeFiHack, \sysname\ (87.04\%) outperformed Smart-LLaMA-DPO~\cite{ref:smartllamadpo2025} (82.35\%), GPTScan~\cite{sun2024gptscan} (79.68\%), VulnHunt-GPT~\cite{boi2024_vulnhunt_gpt} (76.42\%), and general-purpose models such as DeepSeek-R1~\cite{deepseek_r1} (85.23\%) and GPT-4~\cite{chatgpt} (80.36\%). On Web3Bugs the margin over Smart-LLaMA-DPO grew to 6.22 points. Five-fold cross-validation gave mean accuracies of 87.04\% ($\pm$0.42) on DeFiHack and 87.45\% ($\pm$0.89) on Web3Bugs. Bootstrap confidence intervals supported the differences to the LLM baselines.

\subsection{Component Integration and Ablation}

\begin{figure}[t]
    \centering
    \subfloat[DeFiHack]{\includegraphics[width=0.49\columnwidth]{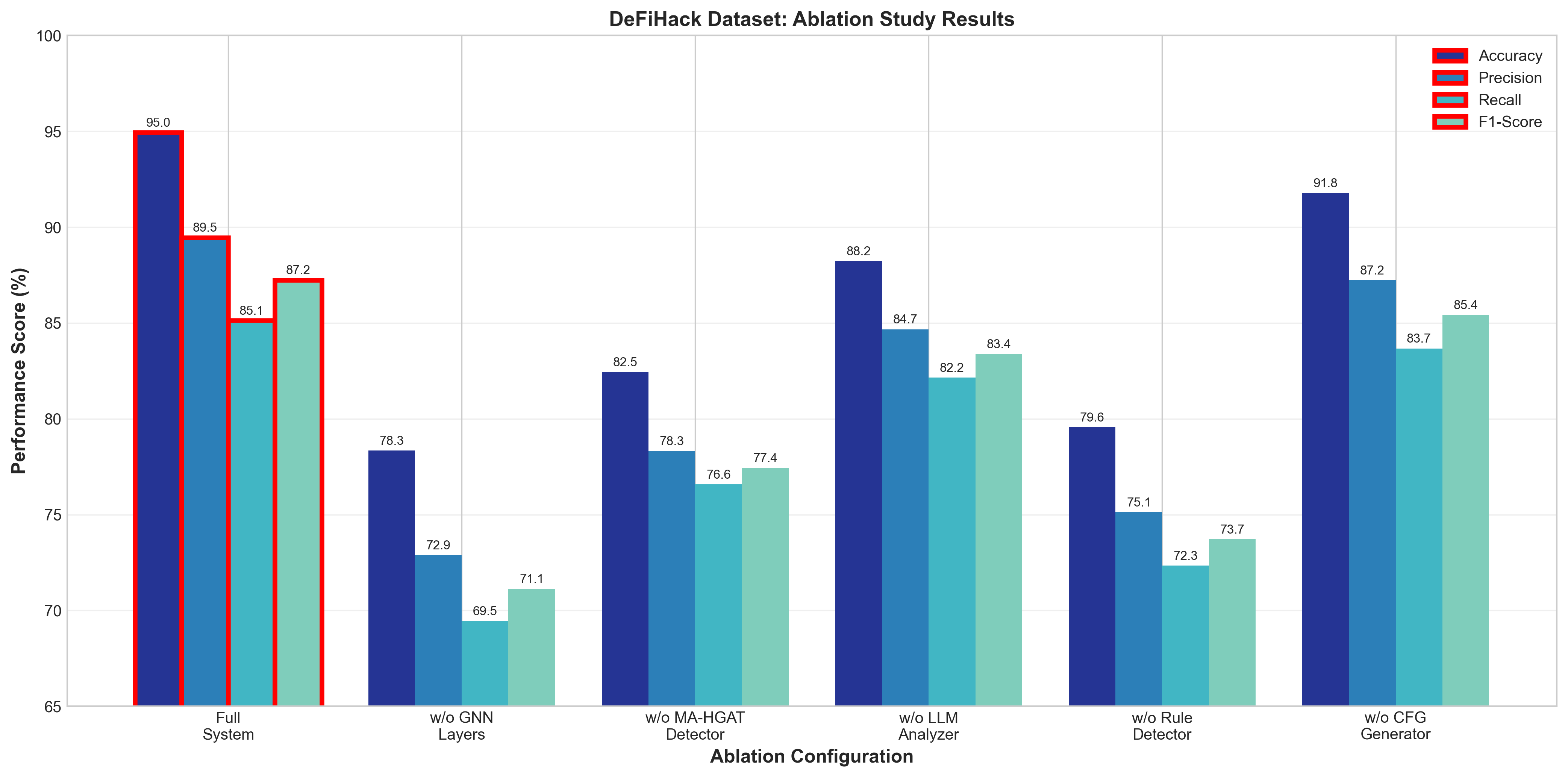}}\hfill
    \subfloat[Web3Bugs]{\includegraphics[width=0.49\columnwidth]{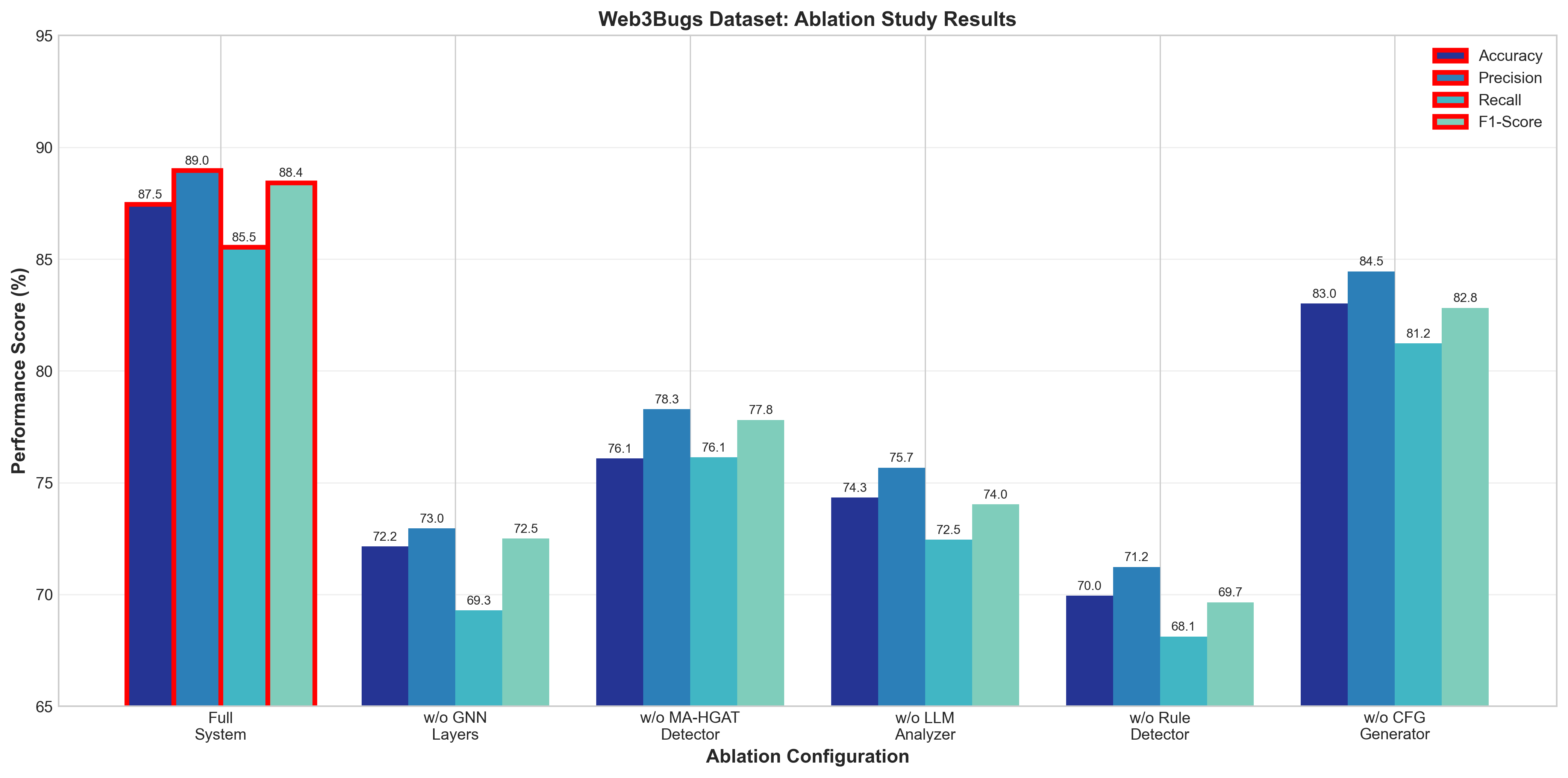}}
    \caption{Effect of removing individual components on the contract datasets.}
    \label{fig:contract_ablation}
\end{figure}

\begin{figure}[t]
    \centering
    \subfloat[DeFiHack]{\includegraphics[width=0.49\columnwidth]{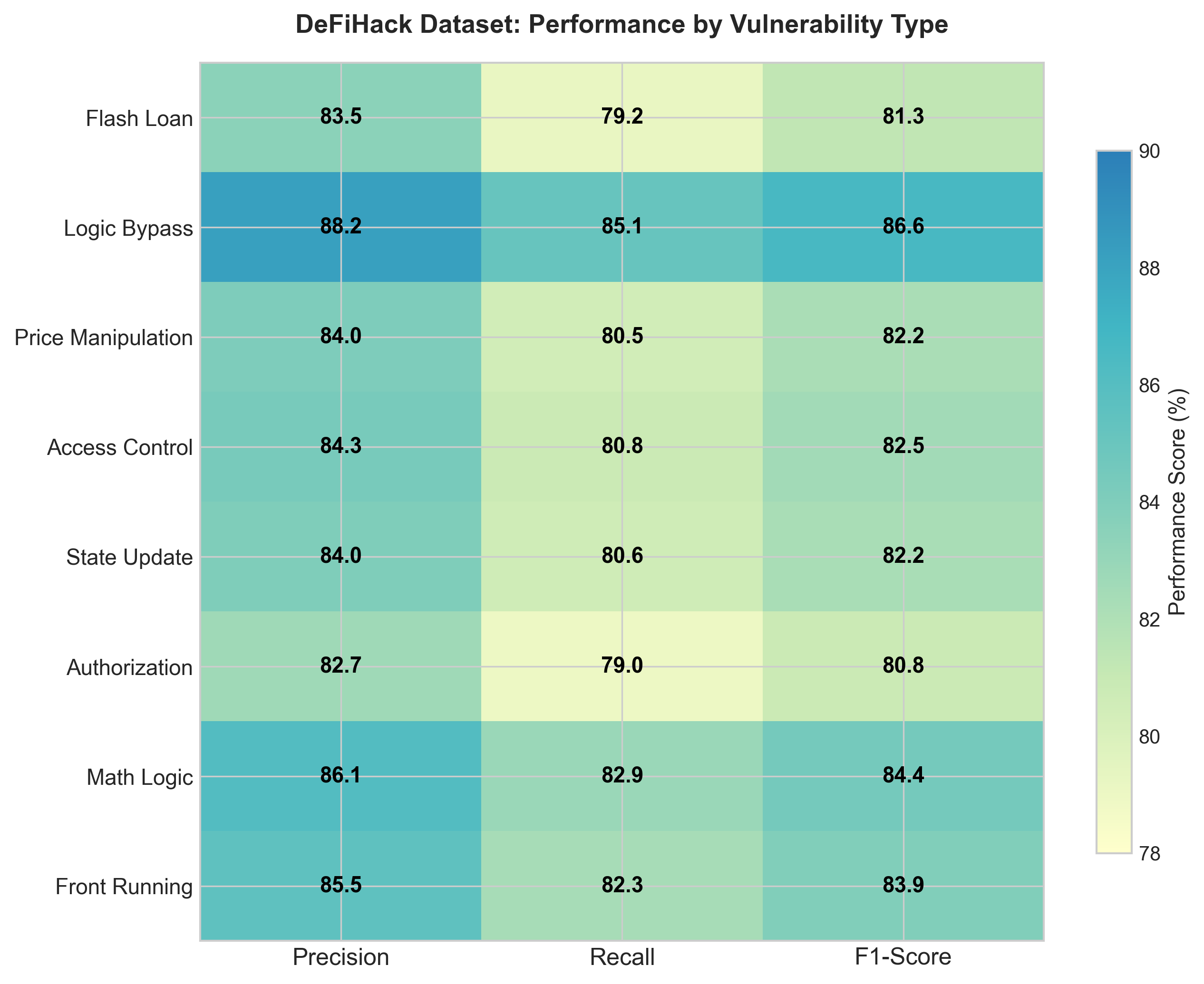}}\hfill
    \subfloat[Web3Bugs]{\includegraphics[width=0.49\columnwidth]{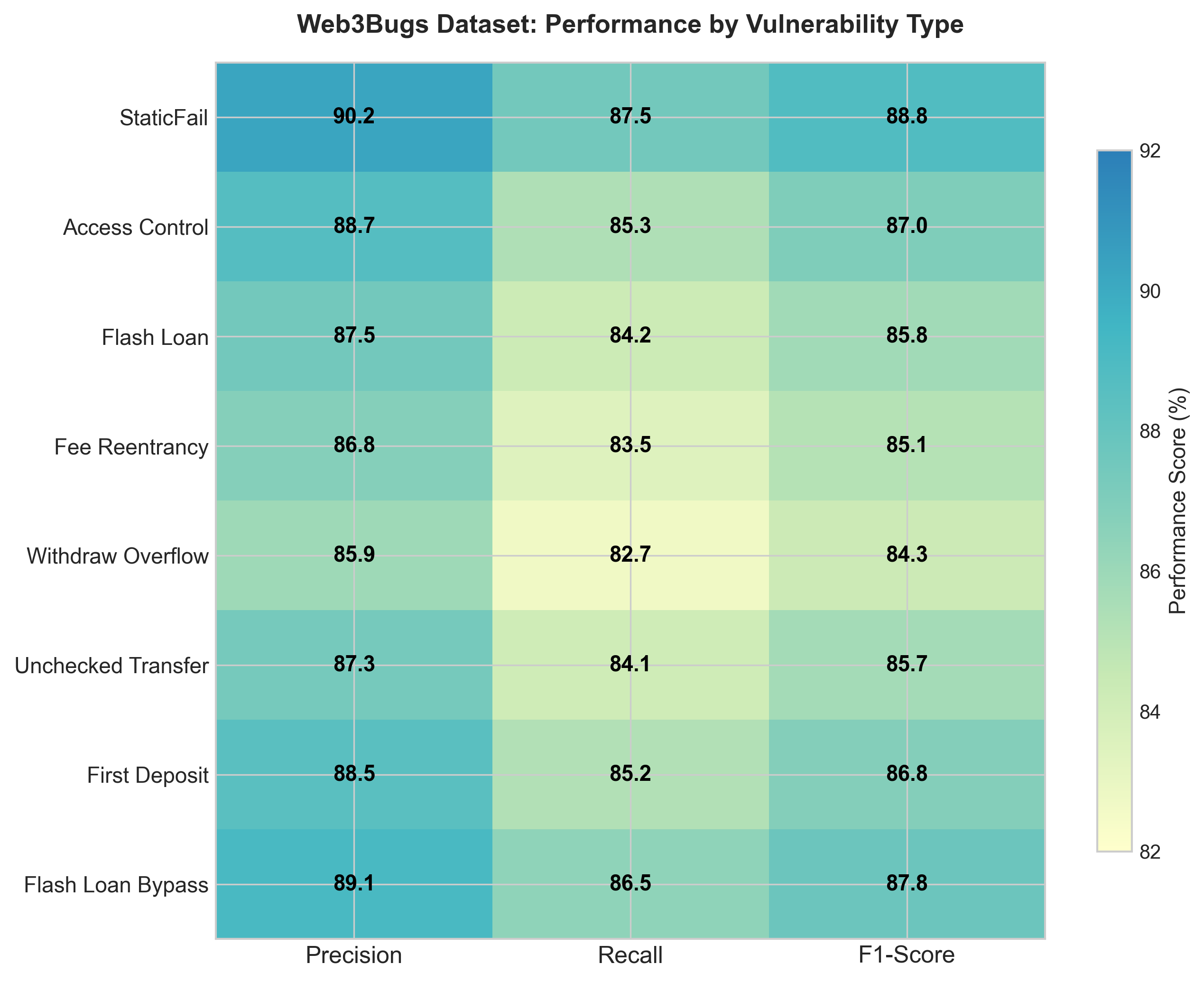}}
    \caption{Per-type detection performance on the contract datasets.}
    \label{fig:contract_vtype}
\end{figure}

\tabref{tab:contract_ablation} reports the ablation of the conference version on DeFiHack. \figref{fig:contract_ablation} shows the effect of removing individual components on both datasets. The full ensemble system combined the graph detector, the rule-based detector, and the LLM analyzer by weighted voting, and it reached 94.95\% accuracy. Within this system, removing the GNN layers caused the largest drop (16.61 points), followed by the rule detector (15.39) and the \sysname\ detector (12.50). Within the graph model, removing the state attention module reduced accuracy by 3.22 points, and removing the multi-agent module reduced it by 4.89. Removing cross-type fusion reduced accuracy by 4.33 points, and collapsing all node types into a homogeneous graph reduced it by 7.11. Per-type analysis showed the strongest results on logic-bypass (86.63\% F1) and mathematical-logic (84.43\% F1) vulnerabilities on DeFiHack. On Web3Bugs, the strongest results were on access-control (90.12\%) and reentrancy (89.45\%) vulnerabilities (\figref{fig:contract_vtype}).

\begin{table}[t]
\centering
\caption{Ablation on DeFiHack as reported in the conference version (\%).}
\label{tab:contract_ablation}
\footnotesize
\begin{tabular}{@{}lcccc@{}}
\toprule
\textbf{Configuration} & \textbf{Acc.} & \textbf{Prec.} & \textbf{Rec.} & \textbf{F1} \\
\midrule
Full system                & \best{94.95} & \best{89.46} & \best{85.13} & \best{87.24} \\
Without GNN layers         & 78.34 & 72.89 & 69.45 & 71.13 \\
Without \sysname\ detector & 82.45 & 78.32 & 76.58 & 77.44 \\
Without LLM analyzer       & 88.23 & 84.67 & 82.15 & 83.39 \\
Without rule detector      & 79.56 & 75.12 & 72.34 & 73.71 \\
Without CFG generator      & 91.78 & 87.23 & 83.67 & 85.42 \\
LLM-only analysis          & 65.42 & 61.78 & 58.34 & 60.01 \\
Rule-only detection        & 58.93 & 55.67 & 52.89 & 54.25 \\
\bottomrule
\end{tabular}
\end{table}

These results were obtained on contract graphs with tens to hundreds of typed nodes and richly populated relations. They established that heterogeneous relational modeling, state-centric attention, and multi-agent fusion each contribute measurably on such graphs. The next section asks whether the same holds at the device layer and what the gateway deployment costs.

\section{Device-Layer and Deployment Results}
\label{sec:iot_experiments}

We now evaluate tasks B and C at the device layer and the deployment of task D at the gateway layer. The released code (per-graph implementation, payload-only features) produces all numbers in this section. Per-fold and per-seed values, per-sample predictions, and the paired tests are released with it. Four questions guide the evaluation. \textbf{RQ1:} Does \sysname\ triage vulnerability reports and link registered firmware images to disclosed weaknesses at least as well as strong homogeneous, tabular, and rule-based alternatives trained on the same information? \textbf{RQ2:} Which of its components matter at the device layer, and why? \textbf{RQ3:} How much of that accuracy survives when the tested devices come from a vendor absent from training? \textbf{RQ4:} What does the role-aligned gateway--cloud split cost, and does it change decisions?

\subsection{Datasets, Tasks, and Protocol}
\label{sec:iot_datasets}

We evaluate the four tasks on four datasets rather than on one deployment, because no public dataset couples the contracts, the firmware, and the transaction stream of a single blockchain-enabled IoT system. Each dataset is the largest public one we found for its layer.

\lead{IoTVulBench (task B, vulnerability-report triage).} IoTVulBench~\cite{iotvulbench2025} provides reproducible vulnerability environments for consumer routers (TP-Link, D-Link, Tenda, and others), each with a curated advisory and a raw PoC request. We use the 95 CVEs that have both. We derive the multi-label target from the advisory tags and text only: command injection (43 CVEs), buffer overflow (43), denial of service (60), and remote code execution (25). Every CVE carries at least one label, and 60 carry exactly two. We build the graphs as in \secref{sec:iot_graphs} from the payload only. Each graph has two function nodes, one to fifteen state nodes (2.3 on average), one or two modifier nodes (only five requests are unauthenticated), and one event node. Two thirds of the graphs have a single state node; for 19 requests that node is a placeholder because no parameter could be parsed. We use five-fold cross-validation stratified on the label combination. Within each fold, we average the predicted probabilities of three random seeds before thresholding at 0.5. We report sample-wise accuracy (fraction of correct label decisions), macro-F1 over the four labels, exact-match rate, and per-label F1. Standard deviations are over folds. As a secondary check, we use a second target on the same graphs: whether the advisory rates the CVE critical (CVSS $\ge 9.0$; 47 of 95).

\lead{FirmVulLinker (task C, fleet linking).} The fleet graph of \secref{sec:iot_graphs} covers 54 images (46 TP-Link, 8 D-Link), 74 CVE groups, and 24 device families. Each CVE affects 11.3 images on average beyond its baseline (1 to 34). For each CVE, we keep its baseline image and a random 70\% of the affected images as known training links. We hold out the remaining 30\% (at least one image) as test links. At inference, \eqnref{eq:link} ranks all images except the baseline and the known positives. We report mean reciprocal rank (MRR), Hits@$K$ for $K\in\{1,3,5,10\}$, and the area under the ROC curve (AUC) per CVE, averaged over CVEs and over three random splits and seeds.

\lead{EdgeChainGuard (task D, deployment workload).} We use the 500-transaction graph (500 function, 3 state, 50 modifier, and 20 event nodes) to measure the gateway--cloud split in \secref{sec:iot_edgecloud}. Its generator assigns attack subtypes stochastically, and its binary attack label coincides with transaction failure~\cite{reis2025edgechainguard}. Every detector, including simple rules and random forests, therefore stays within a few points of the 53\% majority rate on it. We do not report it as a detection result.

\lead{Hyperparameters.} For IoTVulBench: hidden size 64, four heads, dropout 0.25, AdamW with learning rate $10^{-3}$ and weight decay $10^{-3}$, 150 full-batch epochs, per-label binary cross-entropy. For FirmVulLinker: hidden size 128, four heads, dropout 0.2, AdamW with learning rate $2\times10^{-3}$ and weight decay $10^{-4}$, 300 epochs, temperature $\kappa=10$, three negatives per positive. We tuned no hyperparameter on test data.

\lead{Baselines.} The graph baselines receive exactly the same graphs as \sysname. The tabular and rule baselines receive payload statistics only. \emph{Homogeneous GraphSAGE} (labeled \emph{Homogeneous GCN} in the released code) and \emph{Homogeneous GAT} collapse the heterogeneous graph into a single node and edge type. Features are zero-padded or truncated to a common dimensionality. They use two layers of mean aggregation~\cite{hamilton2017graphsage} or graph attention~\cite{velickovic2018gat} with the same hidden size, optimizer, and epochs, followed by mean pooling for graph-level tasks. In the linking task they use the same scorer and negative sampling. The \emph{random forest} operates on hand-crafted graph-level statistics. For PoC traces, these are twelve request-level statistics such as payload length, maximum parameter length and run, entropy, metacharacter count, and endpoint indicators. For linking, they are the concatenated and differenced profiles of the candidate and the baseline image, together with vendor and family agreement. The \emph{rule-based detector} flags injection when shell metacharacters or command tokens appear. It flags overflow when a parameter reaches 200 characters or contains a run of 150 identical characters. It flags denial of service when such a run appears or the request mentions \texttt{reboot} or \texttt{ping}. It flags code execution when injection indicators or such a run are present. For linking we add two more baselines. A \emph{cosine profile} baseline ranks images by profile similarity to the CVE's baseline image. A \emph{metadata rule} ranks images of the same family first and of the same vendor second. Ablations disable the state attention, the multi-agent module, or the cross-type fusion.

\lead{Statistical tests.} For task B we compare methods with a paired bootstrap over the 95 CVEs (10,000 resamples of the macro-F1 difference) and McNemar's test on exact-match correctness. For task C we use a paired bootstrap and a Wilcoxon signed-rank test over the 74 per-CVE reciprocal ranks (averaged over the three splits). We call a difference significant when the 95\% bootstrap interval excludes zero. For the rank-based linking comparisons, whose per-CVE differences are heavy-tailed, we also call it significant when the Wilcoxon test gives $p<0.05$. We report both statistics wherever they disagree.

\subsection{RQ1, Task B: Vulnerability-Report Triage on IoTVulBench}
\label{sec:iot_typing}

\begin{table*}[t]
\centering
\caption{Task B: multi-label vulnerability-report triage on IoTVulBench (95 CVEs, five-fold CV; GNNs average three seeds per fold; \%). Best value per column in bold, second best underlined. Acc.\ is sample-wise accuracy over the four label decisions. Exact is the fraction of CVEs whose entire label vector is correct. CI = command injection, BO = buffer overflow, DoS = denial of service, RCE = remote code execution.}
\label{tab:iotvul}
\footnotesize
\begin{tabular}{@{}lccccccc@{}}
\toprule
\textbf{Method} & \textbf{Acc.\up} & \textbf{Macro-F1\up} & \textbf{Exact\up} & \textbf{F1 CI} & \textbf{F1 BO} & \textbf{F1 DoS} & \textbf{F1 RCE} \\
\midrule
\sysname\ (full)                     & 81.6$\pm$3.5 & \second{78.1$\pm$4.4} & \best{55.8} & 82.5 & \second{90.7} & \second{79.7} & 59.6 \\
~~w/o state attention                & 81.1$\pm$2.6 & 77.7$\pm$3.6 & \best{55.8} & 77.9 & \second{90.7} & \second{79.7} & \best{62.5} \\
~~w/o multi-agent module             & 80.8$\pm$3.5 & 78.0$\pm$4.7 & 52.6 & \second{85.1} & 89.4 & 76.6 & \second{60.9} \\
~~w/o cross-type fusion              & 81.3$\pm$3.3 & 77.9$\pm$3.8 & \second{54.7} & 82.5 & \second{90.7} & 78.9 & 59.6 \\
\midrule
Homogeneous GraphSAGE                & \second{82.6$\pm$1.5} & 77.6$\pm$1.7 & \best{55.8} & 84.7 & 89.4 & \best{80.6} & 55.8 \\
Homogeneous GAT                      & 80.0$\pm$0.5 & 76.1$\pm$2.6 & 46.3 & 83.0 & \second{90.7} & 76.9 & 53.7 \\
Random forest (payload statistics)   & \best{83.9$\pm$2.9} & \best{78.7$\pm$5.8} & 52.6 & \best{87.4} & \best{95.3} & 78.4 & 53.7 \\
Rule-based detector                  & 63.2 & 68.3 & 9.5 & 83.7 & 74.3 & 66.2 & 49.0 \\
\bottomrule
\end{tabular}
\end{table*}

\begin{figure*}[t]
    \centering
    \includegraphics[width=0.98\textwidth]{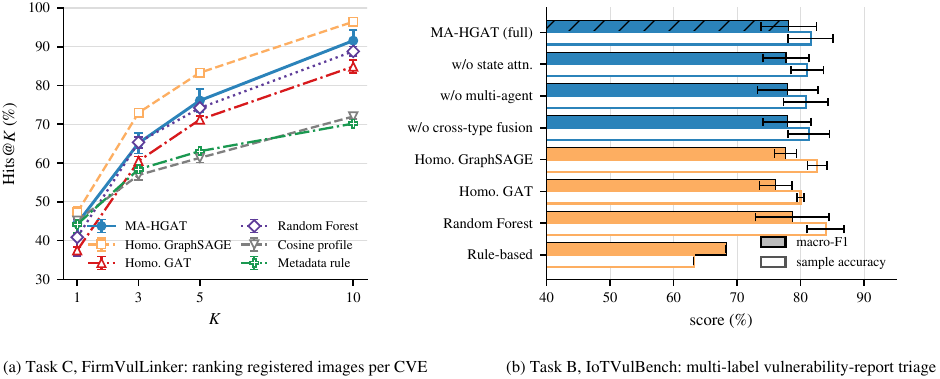}
    \caption{Device-layer accuracy results. (a) Hits@$K$ on FirmVulLinker for the six main methods; error bars are standard deviations over three splits. (b) Macro-F1 (filled) and sample accuracy (hollow) on IoTVulBench; error bars are standard deviations over five folds. Blue bars are \sysname\ variants, orange bars are baselines; the hatched bar is the full model.}
    \label{fig:iot_results}
\end{figure*}

\tabref{tab:iotvul} and \figref{fig:iot_results}(b) report the results. \sysname\ reaches 81.6\% sample accuracy, 78.1\% macro-F1, and 55.8\% exact match. It is clearly better than the rule-based detector. The macro-F1 difference of 9.8 points has a 95\% interval of $[3.4, 16.1]$. McNemar's test on exact matches counts 52 CVEs that \sysname\ labels completely right and the rules do not, against 8 in the other direction ($p<0.001$). The rules fire often (89.8\% recall at 57.5\% precision) and label only 9.5\% of the CVEs completely right. Against the learned baselines, \sysname\ has the highest macro-F1 among the graph models and the highest exact-match rate (tied with GraphSAGE). None of these differences is significant. The macro-F1 difference is $+0.5$ points against GraphSAGE (interval $[-3.0, 4.4]$) and $-0.6$ points against the random forest ($[-5.8, 5.1]$). The exact-match comparison against GraphSAGE is a 7-to-7 tie. The only learned baseline that \sysname\ beats on the paired test is the homogeneous GAT, on exact match (11 CVEs to 2, $p=0.027$). The random forest is the strongest model for buffer overflow (95.3\% F1). There, a single scalar, the longest parameter run, probably decides the label. Tied with the homogeneous GAT, it is also the weakest learned model for remote code execution (53.7\%). \sysname\ and its variants are the strongest there (59.6--62.5\%). For this label, we would expect that evidence from the parameter content, the authentication gate, and the endpoint must be combined.

On the secondary severity target, all learned models separate critical from non-critical advisories well above the 50.5\% majority rate from payload-only graphs. \sysname\ reaches 71.6\% ($\pm$7.1) accuracy and 76.1\% F1, GraphSAGE 75.8\% and 78.9\%, GAT 73.7\% and 77.9\%, and the random forest 72.6\% and 72.3\%. The fold-level standard deviations (6--8 points on 19 CVEs per fold) exceed every pairwise difference.

Our reading is that typed attention has little to attend over on request graphs with five to nineteen nodes. One third of the graphs have exactly five nodes, and two thirds have a single state node. The information that separates the four classes appears to be concentrated in a few parameter statistics that every method receives. A mean aggregator recovers it as well as relation-specific attention does. \sysname's higher fold-to-fold variance ($\pm$4.4 versus $\pm$1.7 for GraphSAGE) is consistent with a higher-capacity model trained on 76 graphs per fold. The graphs are this small because we build them from the request alone. Recovering the handler that parses the request, with its control and data flow, as Karonte~\cite{redini2020karonte} and SaTC~\cite{chen2021satc} do, would populate the schema with far more nodes and relations. The contract results indicate that this is the density at which typed attention starts to pay.

\subsection{RQ1, Task C: Fleet Linking on FirmVulLinker}
\label{sec:iot_linking}

\begin{table*}[t]
\centering
\caption{Task C: linking registered firmware images to disclosed CVEs on FirmVulLinker (54 images, 74 CVE groups; mean$\pm$std over three random splits and seeds). Hits@$K$ in \%. Best value per column in bold, second best underlined.}
\label{tab:firmvul}
\footnotesize
\begin{tabular}{@{}lcccccc@{}}
\toprule
\textbf{Method} & \textbf{MRR\up} & \textbf{Hits@1\up} & \textbf{Hits@3\up} & \textbf{Hits@5\up} & \textbf{Hits@10\up} & \textbf{AUC\up} \\
\midrule
\sysname\ (full)                    & 0.586$\pm$0.027 & 44.4$\pm$3.5 & 65.2$\pm$2.7 & 76.1$\pm$3.0 & 91.6$\pm$2.7 & 0.960$\pm$0.012 \\
~~w/o state attention               & 0.598$\pm$0.015 & 45.7$\pm$2.2 & 66.6$\pm$1.3 & \second{77.2$\pm$1.0} & \second{93.7$\pm$1.4} & 0.966$\pm$0.005 \\
~~w/o multi-agent module            & 0.574$\pm$0.018 & 43.7$\pm$2.5 & 62.8$\pm$2.0 & 74.7$\pm$1.5 & 91.8$\pm$1.6 & 0.957$\pm$0.008 \\
~~w/o cross-type fusion             & \second{0.602$\pm$0.012} & \second{46.1$\pm$2.4} & \second{68.0$\pm$0.5} & 77.1$\pm$2.0 & 93.1$\pm$1.8 & \second{0.967$\pm$0.006} \\
\midrule
Homogeneous GraphSAGE               & \best{0.628$\pm$0.011} & \best{47.3$\pm$1.5} & \best{73.0$\pm$1.0} & \best{83.3$\pm$0.5} & \best{96.5$\pm$0.3} & \best{0.976$\pm$0.003} \\
Homogeneous GAT                     & 0.529$\pm$0.009 & 37.5$\pm$0.8 & 60.5$\pm$1.2 & 71.2$\pm$0.9 & 84.9$\pm$1.7 & 0.916$\pm$0.004 \\
Random forest (pairwise profiles)   & 0.566$\pm$0.021 & 40.9$\pm$4.8 & 65.4$\pm$1.4 & 74.2$\pm$0.5 & 88.8$\pm$1.2 & 0.946$\pm$0.008 \\
Cosine similarity of profiles       & 0.541$\pm$0.010 & 45.2$\pm$1.2 & 56.9$\pm$1.4 & 61.4$\pm$1.2 & 72.0$\pm$0.6 & 0.797$\pm$0.003 \\
Vendor/family metadata rule         & 0.537$\pm$0.012 & 44.2$\pm$2.2 & 58.4$\pm$0.7 & 63.1$\pm$0.1 & 70.2$\pm$0.2 & 0.836$\pm$0.003 \\
\bottomrule
\end{tabular}
\end{table*}

\tabref{tab:firmvul} and \figref{fig:iot_results}(a) report the linking results. Learning over the fleet graph is clearly better than the non-graph heuristics on which vendor advisories implicitly rely. \sysname\ reaches 0.586 MRR, 91.6\% Hits@10, and 0.960 AUC. Raw profile similarity reaches 0.541 MRR and 72.0\% Hits@10, and the vendor/family rule 0.537 MRR and 70.2\% Hits@10. The per-CVE MRR differences of $+0.045$ and $+0.049$ are significant by the Wilcoxon test ($p<0.002$). A few CVEs on which \sysname\ ranks poorly, however, widen the bootstrap intervals to include zero. When an operator checks the top ten candidates per CVE, 92\% of the held-out affected images appear in that list on average with \sysname. With the heuristics, 70--72\% do. The pairwise random forest is between them (0.566 MRR, 88.8\% Hits@10; the difference to \sysname\ is not significant). The homogeneous GAT is clearly worse (0.529 MRR, $p<0.001$).

The homogeneous GraphSAGE baseline, however, is the best model on this task on every metric (0.628 MRR, 96.5\% Hits@10, 0.976 AUC). Its advantage over \sysname\ of 0.042 MRR is close to the significance threshold (bootstrap interval $[-0.088, -0.002]$, $p=0.035$; Wilcoxon $p=0.073$). \sysname\ is also less stable across splits ($\pm$0.027 MRR versus $\pm$0.011). The fleet graph has only two modifier entities (vendors) and 24 event entities (families), and CVE nodes carry a single attribute. All discriminative content is in the 37-dimensional firmware profile of the function nodes and in the $k$-NN and CVE links between them. A mean aggregator over the collapsed graph therefore loses little. Attention that must be learned from 54 images and 74 CVEs probably adds variance rather than signal. We attribute the gap over the heuristics to the graph structure itself, which both graph models exploit. The \textsf{depends} links to known affected images can propagate CVE membership across the $k$-NN \textsf{calls} edges and the shared vendor and family nodes. This is the code-reuse structure that makes vulnerabilities recur~\cite{xiao2024firmrec,cheng2025firmvullinker}.

\subsection{RQ2: Ablation Across Tasks}
\label{sec:iot_ablation}

The ablation rows of \tabref{tab:iotvul} and \tabref{tab:firmvul}, together with the paired tests, give a consistent picture of which components matter at the device layer.

\lead{Only the multi-agent module helps on both tasks.} On IoTVulBench, removing it lowers the exact-match rate from 55.8\% to 52.6\% (macro-F1 changes by 0.1 points). On FirmVulLinker, removing it lowers MRR from 0.586 to 0.574 and Hits@3 from 65.2\% to 62.8\%. The per-CVE reciprocal ranks are higher with the module on 22 CVEs and lower on 7, with 45 ties (Wilcoxon $p=0.005$). The bootstrap interval is $[-0.001, 0.023]$, so the effect does not meet our interval criterion. The effect is small, but it is the one that transfers. We think the per-role refinement and the gated exchange between agents let the sparse vendor and family nodes influence the image embeddings without being averaged away.

\lead{Cross-type fusion is neutral or slightly harmful.} On IoTVulBench, removing it changes macro-F1 by 0.2 points and exact match by 1.1 points (neither significant). On FirmVulLinker, removing it raises MRR from 0.586 to 0.602 (not significant, $p=0.19$). The fusion layer re-injects graph-level context. In a single fleet graph that context is identical for every pair and only adds parameters.

\lead{State attention helps only when states carry structure.} On IoTVulBench, state nodes are HTTP parameters with ten attributes each, and their order matters. There, removing the KVQ module costs 0.4 macro-F1 points and 4.6 F1 points on command injection, the label most tied to parameter content. Exact match is unchanged. On FirmVulLinker, a state node is a CVE with a single year attribute. There, removing the module slightly \emph{improves} MRR (0.598 versus 0.586, $p=0.10$) and Hits@10 (93.7\% versus 91.6\%). We attribute this to the module's positional encoding and softmax over 74 near-identical CVE embeddings, which have no structure to exploit. This mirrors, in reverse, the conference finding that state attention was worth 3.22 accuracy points on contracts, where state variables are the richest role.

\lead{Takeaway.} Typed heterogeneity pays off when (i) several roles carry informative attributes and (ii) the graphs are large enough for relation-specific attention to be estimated. Contract graphs satisfy both. The two device-layer benchmarks available today satisfy neither, and there a mean aggregator over the same graph is at least as good. At the device layer, the value of the schema is not higher accuracy. It lies in representing both layers of the system with one model and in the deployment property of \secref{sec:iot_edgecloud}.

\subsection{RQ3: Generalization to an Unseen Vendor}
\label{sec:crossvendor}

The protocols above draw training and test cases from the same pool of devices, so we reran both tasks under a vendor holdout. For task B we resolve each CVE's vendor from the advisory text alone (54 D-Link, 24 Tenda, 17 TP-Link), leave one vendor out at a time, and pool the three folds. For task C the CVE groups are vendor-pure, so we train on one vendor's groups and hold out every link of the other's except the query image that defines each CVE. Everything else stays as in \secref{sec:iot_datasets}.

\begin{table}[t]
\centering
\caption{Vendor holdout. Task B reports macro-F1, task C mean reciprocal rank, both in \%. ``In dist.'' repeats \tabref{tab:iotvul} and \tabref{tab:firmvul}. The heuristic row is the rule-based detector for task B and the vendor/family rule for task C; the random forest uses payload statistics for task B and pairwise profiles for task C.}
\label{tab:crossvendor}
\footnotesize
\setlength{\tabcolsep}{3.5pt}
\begin{tabular}{@{}lcccccc@{}}
\toprule
 & \multicolumn{2}{c}{\textbf{Task B, F1}} & \multicolumn{3}{c}{\textbf{Task C, MRR}} \\
\cmidrule(r){2-3}\cmidrule(l){4-6}
\textbf{Method} & \textbf{In dist.} & \textbf{Unseen} & \textbf{In dist.} & \textbf{TP$\rightarrow$D} & \textbf{D$\rightarrow$TP} \\
\midrule
\sysname\ (full)        & 78.1 & 59.4 & 58.6 & 51.6 & 16.5 \\
Homogeneous GraphSAGE   & 77.6 & 55.3 & 62.8 & 56.2 & \second{16.7} \\
Homogeneous GAT         & 76.1 & \second{60.5} & 52.9 & 55.6 & 13.4 \\
Random forest           & 78.7 & 53.6 & 56.6 & 3.9  & 8.5 \\
Heuristic (no training) & 68.3 & \best{68.3} & 53.7 & \best{63.4} & 14.0 \\
Cosine profiles         & --   & --   & 54.1 & \second{56.4} & \best{20.2} \\
\midrule
Random ranking          & --   & --   & --   & 8.7  & 8.7 \\
\bottomrule
\end{tabular}
\end{table}

\lead{Neither task transfers across vendors.} On task B every learned model loses 19 to 25 macro-F1 points, and exact match falls from 55.8\% to 21.1\% for \sysname. No difference between \sysname\ and any baseline is significant on the paired bootstrap, and the widest interval is $[-10.3, 1.8]$ points against the random forest. The rule-based detector does not train, so its row is unchanged, and under the shift it has the highest macro-F1 of all. On task C the two directions disagree because of the corpus composition. Only 8 of the 54 images are D-Link, so testing on D-Link rewards anything that ranks the query's vendor first, and the metadata rule beats \sysname\ there (0.634 against 0.516, interval $[-0.21, -0.04]$). The reverse direction removes that shortcut. Every method then falls near profile similarity and far below the 0.586 of the in-distribution split.

\lead{Reading.} Both benchmarks describe a device through vendor-specific surfaces. These are HTTP endpoints and parameter names for task B, and firmware layout for task C. Under a vendor shift the evidence a model relies on changes, and the learned mapping does not follow it. Inside the device population it was trained on, the framework is competitive with strong baselines. Outside it, an operator should retrain rather than transfer. Carrying attributes that describe what a handler does, rather than where it sits, is the concrete next step for the schema.

\subsection{RQ4, Task D: Gateway--Cloud Deployment}
\label{sec:iot_edgecloud}

\begin{figure*}[t]
    \centering
    \includegraphics[width=0.98\textwidth]{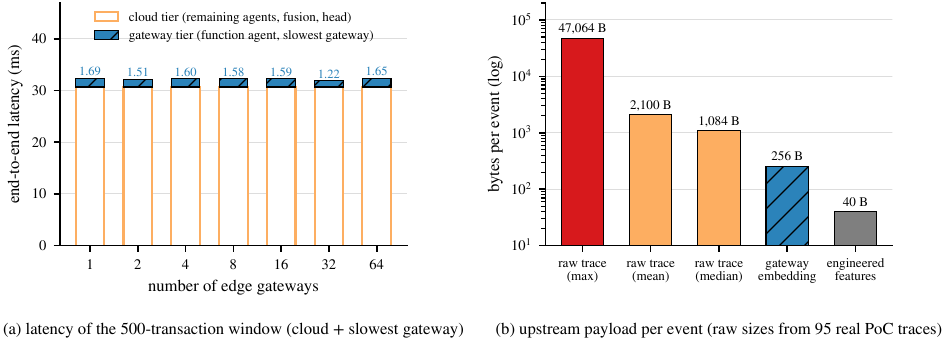}
    \caption{Cost of the gateway--cloud split. (a) End-to-end latency for the 500-transaction window as the window is partitioned over 1--64 gateways. The cloud tier (hollow, 30.67~ms) is independent of the number of gateways. The gateway tier (hatched, numbers above bars) is the latency of the slowest gateway. (b) Upstream payload per event: the 256-byte gateway embedding versus the raw PoC requests of IoTVulBench (mean, median, maximum) and a pre-extracted 40-byte feature vector.}
    \label{fig:edge_scaling}
\end{figure*}

We measure the split of \secref{sec:edgecloud} on the EdgeChainGuard stream (500 transactions, 50 devices) with the IoTVulBench PoC requests as the reference for raw request sizes. All measurements are CPU inference in PyTorch with intra-op threads pinned to one, 10 warm-up and 50 timed repetitions (20 for the per-gateway timings of \figref{fig:edge_scaling}(a)). The timed cloud-only forward pass corresponds to the full model of \secref{sec:architecture} on the whole window. The gateway-tier figure in \tabref{tab:edgecloud} includes the construction of the gateway's local graph, whereas the per-gateway figures in \figref{fig:edge_scaling}(a) time the forward pass only. \tabref{tab:edgecloud} and \figref{fig:edge_scaling} summarize the results.

\begin{table}[t]
\centering
\caption{Gateway--cloud partitioning of \sysname\ on the 500-transaction EdgeChainGuard window (CPU inference, one thread).}
\label{tab:edgecloud}
\footnotesize
\begin{threeparttable}
\setlength{\tabcolsep}{4pt}
\begin{tabular}{@{}lrr@{}}
\toprule
\textbf{Quantity} & \textbf{Cloud-only} & \textbf{Split} \\
\midrule
Parameters at the gateway (of 204{,}802) & 0     & 5{,}120 (2.5\%) \\
Parameter footprint, gateway (KB)   & --       & 20.5 \\
Parameter footprint, cloud (KB)     & 819.2    & 798.7 \\
Activation bytes, gateway (KB)\tnote{a} & --   & 1{,}038 \\
Activation bytes, cloud (KB)\tnote{a}   & 8{,}983 & 8{,}983 \\
Embedding buffer per window, gateway (KB) & -- & 128 \\
Latency, gateway tier (ms)          & --       & 2.04 \\
Latency, cloud tier (ms)            & 31.22    & 30.67 \\
Latency, end to end (ms)            & 31.22    & 31.66 \\
Uplink per event (B)\tnote{b}       & 2{,}100  & 256 \\
Uplink per 500-event window (KB)    & 1{,}050  & 128 \\
Accuracy (\%)\tnote{c}              & 57.5$\pm$0.7 & 56.5$\pm$1.1 \\
Macro-F1 (\%)\tnote{c}              & 57.0$\pm$1.1 & 55.8$\pm$1.1 \\
\bottomrule
\end{tabular}
\begin{tablenotes}\footnotesize
\item[a] Sum of the sizes of all tensors emitted by the modules during one forward pass over the window, an upper bound on the live activation working set. The cloud figure is for the full model.
\item[b] Cloud-only uploads the raw request; the value is the mean of the 95 IoTVulBench PoC requests (median 1{,}084~B, maximum 47{,}064~B). The split uploads one 64-dimensional embedding. A pre-extracted 10-dimensional feature vector would be 40~B.
\item[c] Mean$\pm$std over three seeds on a held-out 200-transaction split (majority rate 50.5\%; 300 training transactions, unweighted cross-entropy, 120 epochs). The split variant is trained end to end with the full split pipeline, including the fusion layer. Both values are near chance because the workload's labels are synthetic. The row is a parity check, not a detection result.
\end{tablenotes}
\end{threeparttable}
\end{table}

\lead{The gateway model is tiny.} The Function agent's gateway share (the function projection and the \textsf{calls} attention block) has 5,120 parameters, 2.5\% of the 204,802-parameter model. That is 20.5~KB in single precision versus 819~KB for the full model. Its activations for a 500-event window sum to at most 1.04~MB (2.1~KB per event), one ninth of the 8.98~MB of the full model. Its output buffer is 128~KB (500 embeddings of 64 floats). A footprint of this size fits the gateway-class devices that relay device transactions.

\lead{The split is almost free in latency.} Cloud-only inference over the window takes 31.22~ms. The split takes 2.04~ms at the gateway (including the construction of the local graph) and 30.67~ms in the cloud. Measured end to end, it takes 31.66~ms, within 1.5\% of the cloud-only figure. The conservative sum of the gateway and cloud tiers is 32.7~ms, within 5\%. As \figref{fig:edge_scaling}(a) shows, the cloud time does not change with the number of gateways. The time of the slowest gateway stays between 1.2 and 1.7~ms from 1 to 64 gateways. The end-to-end latency of a window therefore stays between 31.9 and 32.4~ms for any partition, and a single gateway sustains about $3.3\times10^{5}$ events per second. These times come from a general-purpose CPU pinned to one thread, not from gateway hardware, so they transfer as ratios and not as absolute values. The parameter and activation footprints above transfer directly.

\lead{Raw requests stay at the gateway.} A gateway uploads a 256-byte embedding per event instead of the raw request. In IoTVulBench, a raw request averages 2,100~B (median 1,084~B, maximum 47,064~B), so the reduction is 8.2$\times$ on average and 184$\times$ in the worst case. The embedding is larger than a pre-extracted 40-byte feature vector, so the benefit over feature-level offloading is not bandwidth. Instead, the gateway does not need to run and update a domain-specific feature extractor. Request contents, which may embed credentials or configuration values, also never leave the device side.

\lead{Decisions do not change.} Over three seeds, the model trained end to end with the split in place, including the fusion layer, reaches 56.5\% ($\pm$1.1) accuracy and 55.8\% ($\pm$1.1) macro-F1. Cloud-only training reaches 57.5\% ($\pm$0.7) and 57.0\% ($\pm$1.1). The difference of about one point is comparable to the seed-to-seed spread. Both values are near chance because this workload's labels are synthetic (\secref{sec:iot_datasets}). The row is a parity check, not a detection result.

\section{Conclusion}
\label{sec:conclusion}

Blockchain-enabled Internet of Things systems rely on security-critical logic in both smart contracts and device firmware, yet existing approaches typically analyze these layers separately. This work addressed this gap by extending \sysname\ into a unified cross-layer framework that represents contracts, firmware evidence, device fleets, and device--contract transactions through the same role-based structure. The framework supports contract auditing, vulnerability-report triage, fleet-level vulnerability linking, and transaction monitoring within one model, while allowing the device-action component to run on resource-constrained gateways.

Our evaluation shows that \sysname\ is effective across both contract and device layers and that the role-aligned multi-agent design is the main source of cross-layer transfer. The gateway--cloud partition further reduces local computation and avoids transmitting raw requests with little impact on decision latency. At the same time, the unseen-vendor evaluation reveals a clear generalization limitation. Future work will focus on richer firmware-derived representations, broader labeled datasets, and end-to-end evaluation on real deployments.

\bibliographystyle{IEEEtran}
\bibliography{references_iotj}






\end{document}